\documentclass[11pt]{article}

\usepackage{osid}

\usepackage{amsmath}
\usepackage{dsfont}
\usepackage{hyperref}
\usepackage{todonotes}
\DeclareMathOperator{\tr}{Tr}

\newcommand{\id}{\mathds{1}}

\renewcommand{\i}{\mathrm i}
\renewcommand{\L}{\mathrm L}
\renewcommand{\L}{\mathcal{G}}

\newcommand{\e}{\mathrm e}
\renewcommand{\eqref}[1]{Eq.~(\ref{#1})}

\usepackage[normalem]{ulem}

\newcommand{\map}{\mathcal{E}}

\newcommand{\states}{\mathcal{S}}

\newcommand{\RomanNumeralCaps}[1]
    {\MakeUppercase{\romannumeral #1}}
    \usepackage{float}

\usepackage{svg}
\usepackage{tikz}
\usetikzlibrary{intersections,arrows.meta}
\usepackage[dvipsnames]{xcolor}
\usepackage{cleveref}
\usepackage{subcaption}
\usepackage{graphicx}

\title{Gaussian and Qubit Semigroup Dynamics beyond GKSL}
\author{Nadine Diesel \\{\footnotesize\it Institut für Theoretische Physik, Technische Universität Dresden, D-01062, Dresden, Germany \& nadine.diesel@tu-dresden.de}\\[2ex]
Charlotte Bäcker \\{\footnotesize\it Institut für Theoretische Physik, Technische Universität Dresden, D-01062, Dresden, Germany \& charlotte.baecker@tu-dresden.de}\\[2ex]
Walter T. Strunz \\{\footnotesize\it Institut für Theoretische Physik, Technische Universität Dresden, D-01062, Dresden, Germany  \& walter.strunz@tu-dresden.de}\\[0.2ex]}

\begin{document}

\maketitle

\begin{abstract}
A well established approach to describe open quantum system dynamics is the celebrated Gorini-Kossakowski-Sudarshan-Lindblad (GKSL) master equation. In this article we aim to shed light on semigroup dynamics that are not of this form. Such dynamics may appear in real physical systems describing asymptotic long-time dynamics. Thus, they correspond to semigroup dynamics applicable to a restricted set of initial states only. As a way of motivating our studies, first we review known results about the single-mode Gaussian \emph{Quantum Brownian motion} channel and corresponding master equations. Next we discuss a conceptually simpler family of qubit semigroup generators and identify parameter regions of completely positive, merely positive, and non-positive dynamics. We then focus on the latter case to which we will refer to as \emph{restricted-positive}, as the semigroup generator may well be physically meaningful when applied to a restricted set of admissible initial states.  
\end{abstract}

\section{Introduction}\label{sec:Intro}

Open system quantum dynamics is concerned with the time evolution of the reduced state $\rho(t)$ of a system, ignoring knowledge about the degrees of freedom of the environment. Similar to the Schrödinger equation of a closed system, and being computationally straightforward, one often aims at describing the dynamics through a differential equation of first order in time -- a master equation -- relating $\dot\rho(t)$ to the current state $\rho(t)$ of the open system. 

Thus, for a time independent generator ${\L}$ the master equation reads
\begin{equation}\label{eq:semigroup_master}
    \dot\rho(t) = {\L}[\rho(t)],\;\;
\end{equation}
ensuring that the corresponding propagating map ${\mathcal E}(t)=\e^{\L t}$ naturally satisfies the semigroup property
\begin{equation}\label{eq:semigroup_property}
   {\mathcal E}(t+s)={\mathcal E}(t){\mathcal E}(s), \qquad t,s\geq0.
\end{equation}
Demanding complete positivity of ${\mathcal E}(t)$, we know that the generator has to be of the celebrated Gorini-Kossakowski-Sudarshan-Lindblad (GKSL) form~\cite{Lin1976aRepMathPhys,GorKosSud1976,benatti2026fiftyyearsgklsmaster}
\begin{equation}\label{eq:gksl_master}
    \L[\rho] = -\i [H,\rho] +
      \sum_n \left(L_n\rho L_n^\dagger -\frac{1}{2} \{L_n^\dagger L_n,\rho\}\right),
\end{equation}
where $\{A,B\}=AB+BA$ is the anti-commutator and
$H$ is a Hermitian operator taking the role of the Hamiltonian. The irreversible part of the dynamics is taken care of by an arbitrary number  of \emph{jump operators} $L_n$. We stress that the condition on ${\mathcal E}(t)$ to be a completely positive and trace preserving (CPT) map implies that the initial state of the dynamics can be chosen from the full set $\states$ of possible states of the open system: $\rho(0)\in\states$.

Indeed, as highlighted profoundly in the Pechukas-Alicki exchange about the relevance of CPT maps for open system dynamics \cite{pechukas1994,alicki1995,pechukas1995}, demanding complete positivity is intimately connected to the assumption of an uncorrelated initial state of system and environment. For realistic open system dynamics -- and in particular for strongly damped open systems -- this assumption may be too restrictive. 
Although the assumption of a semigroup dynamics \eqref{eq:semigroup_property} may be physically justified, possible pre-existing correlations between system and environment render the CPT-framework too narrow. This implies that a generator of the form of \eqref{eq:gksl_master} is not mandatory for a master equation formulation to describe a valid physical evolution of a quantum system.

We will elaborate on these issues in the next section based on the established example of the quantum Brownian motion master equation. In essence, non-GKSL semigroup generators can be physically relevant, yet one has to discuss (and restrict) the set of admissible initial states $\rho(0)$ (or lifts thereof), reflecting initial correlations between system and environment \cite{pechukas1994}. Thorough discussions of the role of initial system-environment correlations for the corresponding open quantum dynamics can be found in  \cite{Buzek2001,Sudarshan2003,JordanShajiSudarshanPRA2004,ShajiSudarshanPLA2005,Carteret2008,dominy2015,DominyLidarQIP2016,Alipour2020,ColNeuBre2022,BenattiChruscinskiPascazioArXiv2605.18639}. In \cite{Buzek2001,JordanShajiSudarshanPRA2004,Carteret2008} a general study of the structure of such dynamics is offered, while maps beyond the CPT class are studied in \cite{Sudarshan2003,ShajiSudarshanPLA2005}, with an emphasis on their physical relevance. The latter issue is also investigated in \cite{dominy2015} for general (non-CP and non-positive) maps. For a more explicit study of admissible (restricted) domains for open system initial states under non-CP maps see \cite{JordanShajiSudarshanPRA2004,DominyLidarQIP2016}. Master equation approaches to dynamics with initial system-environment correlations are presented in \cite{Alipour2020,ColNeuBre2022}. For a recent detailed study of open system dynamics with entangled initial states, see \cite{BenattiChruscinskiPascazioArXiv2605.18639}. Those authors offer a criticism of the approach based on restricting the set of initial states, taken in this contribution. They argue that for non-CPT dynamics the volume of the restricted set of admissible initial conditions may become very small which they consider to be a significant indication that complete positivity should be upheld in open quantum system dynamics. In our work, we circumvent an explicit discussion of initially correlated system-environment states. Instead, our approach is largely local by studying directly the (complete) positivity properties of non-GKSL master equations. 

Indeed, it may well happen that one comes across meaningful master equations that are \emph{not} of the desired GKSL form \eqref{eq:gksl_master}. Well-known examples are given by
the second-order perturbative Redfield master equation \cite{Redfield1957,Whitney2008,Hartmann2020, kaplanek2026redfield}, or master equations related to what we can generally call quantum Brownian motion: a quantum particle with canonical position and momentum operators $(\hat q,\hat p)$, coupled linearly to a bath of harmonic oscillators, characterized by a certain bath correlation function \cite{Ullersma1966a,Ullersma1966b,Caldeira1983,HaakeReibold1985,HuPazZhangPRD1992}. We will discuss relevant results about quantum Brownian motion in detail in Secs. \ref{sec:qbm_master} and \ref{sec:qbm_model}. 

Since non-GKSL master equations may violate (complete) positivity, a significant effort has been devoted to {\emph{cure}} them, i.e.\ to bring these non-GKSL generators into GKSL form \cite{Schaller2008,Farina2019,Tupkary2022,Pradilla2024}. This can be achieved either by applying further approximations (most notably the secular or rotating wave approximation), accepting, however, a poorer quality of the ``cured'' master equation \cite{Hartmann2020}. Sometimes, an a priori non-GKSL master equation can be brought into GKSL form by simply adding suitably chosen terms to the generator \cite{Diosi1993} -- these may be justified, again, by applying additional approximations (e.g.\ time coarse graining \cite{Diosi1993,Diosi1993b}).

As outlined above, in this contribution we would like to take the different point of view already taken in \cite{pechukas1994}, and in various forms also in
\cite{Buzek2001,Sudarshan2003,JordanShajiSudarshanPRA2004,ShajiSudarshanPLA2005,Carteret2008,dominy2015,DominyLidarQIP2016,Alipour2020,ColNeuBre2022}: we consider the problems arising from non-GKSL generators to \emph{not} be connected to the generator of the dynamics, but to the set of initial states these generators are supposed to be applied to. Accepting the non-GKSL generators for what they are, their physical relevance emerges from a restriction on the possible initial conditions, ultimately related to pre-existing correlations between system and environment. Leaving the CPT framework for master equations entails fewer restrictions on the semigroup generator. For example, restrictions on the relative strength of decay rates no longer apply \cite{pechukas1994,SchirmerSolomonPRA2004,ChruscinskiKimuraShishidoPRL2021,ChruscinskietalRepProgPhys2025}.

For CPT dynamics, the initial state of system and environment is an uncorrelated state that may well be a highly quenched initial state, considering the system-environment interaction present in the total Hamiltonian. Thus, starting the coupled dynamics from such a quenched product state,
we often observe a very fast transient dynamics to a correlated system-environment state. It is only after this transient ``initial slip'' \cite{Whitney2008, Diosi1993b,HaakeLewensteinPRA1983,Suarez1992,StrunzDiosi1999,   Gaspard1999,StrachanPurkayasthaClarkJChemPhys2024} or ``initial jolt'' \cite{HuPazZhangPRD1992,FlemingHuRouraAoP2011,FlemingRouraHuPRE2011}, that the much slower equilibration dynamics -- governed by a master equation -- becomes relevant. In other words, as this initial, fast transient build-up of correlations is typically on the time scale of bath correlations, a standard master equation (with a time independent generator) will not be able to capture this dynamics. Still, after this initial fast transient regime, the dynamics may well be described by a master equation. Note, however, that in this case the corresponding set of initial states of the master equation is a strict subset of all states due to system-environment correlations. The set is comprised of those states that have already been exposed to a certain level of decoherence during the initial transient slip. These observations are well established in connection with the Brownian motion model, with an
infinite-dimensional Hilbert space \cite{HaakeReibold1985, HuPazZhangPRD1992}. Here, we apply and explore similar concepts for a single qubit.

This paper is structured as follows: We devote the first half to a continuous-variable model in order to review known results and motivate our work on the later qubit channels in the second half.
In Sec.~\ref{sec:qbm_master} we concern ourselves with single-mode Gaussian dynamics and review conditions on their physicality. In Sec.~\ref{sec:qbm_model} the explicit quantum Brownian motion model is considered. After reviewing the exact master equation, we focus on the semigroup equation describing the dynamics after the transient evolution. The concepts of restricted (complete) positivity introduced there are then discussed thoroughly in the context of qubit dynamics in Sec.~\ref{sec:qubit-semigroup}. After proposing a family of suitable non-GKSL generators and their representation, we investigate under which circumstances the resulting dynamics is (completely) positive and under which restricted positive. We finally draw conclusions and discuss implications for the form of master equations describing physical dynamics.

\section{Single-mode Gaussian GKSL Semigroup}\label{sec:qbm_master}

We first consider a one-dimensional continuous variable system with dimensionless position and momentum operator $(q,p)$ such that $[q,p]=\i$. A generator 
$\L=\L(q,p)$, whose propagator $\exp(\L t)$ defines a Gaussian map is at most quadratic in $(q,p)$. In its most general form \cite{Lin1976aRepMathPhys, DekkerValsakumar1984PhysLettA,Sandulescu_1986} it may be written as
\begin{equation}\label{eq:gaussian_master}
    \L[\rho] = -\i [H,\rho] - \i\gamma[q,\{p,\rho\}] + 2D_{pq} [q,[p,\rho]] -D_{qq}[p,[p,\rho]] -D_{pp} [q,[q,\rho]],
\end{equation}
where $H=H(q,p)$ is the Hamiltonian and $\{p,\rho\}=p\rho+\rho p$ the anti-commutator.  In Wigner representation,
the master equation takes the form of a Fokker-Planck equation corresponding to a standard phase-space Ornstein-Uhlenbeck process \cite{DekkerValsakumar1984PhysLettA,Sandulescu_1986}. Then it becomes apparent that the parameter $\gamma>0$ turns out to be the damping rate, while the $D_{ij}$ are diffusion constants. As Hamiltonian, we choose
\begin{equation}\label{eq:Hamiltonian}
    H=\frac{1}{2} p^2 + \frac{\eta}{2}(qp+pq) + \frac{1}{2}\Omega^2 q^2,
\end{equation}
disregarding a possible additional linear term $fq+ vp$. 

\subsection{(Complete) positivity condition}
\label{sec:complete_positivity_condition}
It was shown by various authors \cite{Lin1976aRepMathPhys, Sandulescu_1986} that \eqref{eq:gaussian_master} is a proper GKSL generator, if and only if the diffusion coefficients $D_{ij}$ and damping rate $\gamma$ satisfy the conditions
$D_{qq}>0,\;\;D_{pp}>0\,;\;D_{qq}D_{pp}-D_{pq}^2\ge\frac{\gamma^2}{4}$. These restrictions also appear in the work of Dekker and Valsakumar \cite{DekkerValsakumar1984PhysLettA}. Introducing the diffusion matrix $D=(D_{ij})$ the CP condition can be concisely written as
\begin{equation}\label{eq:dekker-condition}
    D \succ 0,\;\;\;\;\det D\ge\frac{\gamma^2}{4}.
\end{equation}
Here and in the following $D\succ0$ for a matrix means it is positive definite.
With these restrictions on the parameters, we can always construct two jump operators $L_n=\alpha_n q + \beta_n p$ with $\alpha_n,\beta_n\in \mathrm{C},\;\; (n=1,2)$ such that the general quadratic generator from \eqref{eq:gaussian_master} takes the GKSL form of \eqref{eq:gksl_master}. Conversely, any GKSL generator \eqref{eq:gksl_master} with a Hamiltonian at most quadratic in $(q,p)$ and an arbitrary number of at most linear jump operators $L_n=\alpha_n q + \beta_n p$ leads to the form \eqref{eq:gaussian_master} with parameters that satisfy the conditions provided by \eqref{eq:dekker-condition}. Thus, with conditions \eqref{eq:dekker-condition}, the generator \eqref{eq:gaussian_master} is such that $\exp(\L t)$ is a Gaussian quantum channel, i.e.\ a Gaussian CPT map for all $t$.

It is worth noting that in \cite{DekkerValsakumar1984PhysLettA}, Dekker and Valsakumar arrive at the conditions of \eqref{eq:dekker-condition} {\emph{not}} through an analysis of complete positivity: instead, they consider the dynamics of the
variances of the canonical variables,
\begin{equation}\label{eq:covariance}
    \sigma_{qq}=\langle q^2\rangle-\langle q\rangle ^2,\;\;
     \sigma_{qp}=\sigma_{pq}=\frac{1}{2}\langle qp + pq\rangle-\langle q\rangle\langle p \rangle,\;\;\sigma_{pp}=\langle p^2\rangle - \langle p\rangle^2.
\end{equation}
We combine these in the covariance matrix $\sigma$ through
\begin{equation}\label{eq:covariancematrix}
    \sigma = \begin{pmatrix}
        \sigma_{qq} &\sigma_{qp} \\
        \sigma_{qp} & \sigma_{pp} 
    \end{pmatrix}
\end{equation}
A positive matrix as in \eqref{eq:covariancematrix} is the covariance matrix of a quantum state if and only if it satisfies the Robertson-Schrödinger uncertainty relation~\cite{Robertson1929,Schrodinger1930,Heinosaari2010}
\begin{equation}\label{eq:gauss_state_condition}
    \det\sigma\ge\frac{1}{4}.
\end{equation}
Dekker and Valsakumar show in \cite{DekkerValsakumar1984PhysLettA} that if conditions \eqref{eq:dekker-condition} are not met, one can always find Gaussian quantum states with initially $\det\sigma(0)\ge 1/4$ that turn into non-positive operators under time evolution given by \eqref{eq:gaussian_master}, i.e.\ $\sigma(t)$ will violate \eqref{eq:gauss_state_condition} for some time $t>0$. In other words, what they show is that if the Gaussian map induced by the dynamics under \eqref{eq:gaussian_master} is (merely) positive, conditions \eqref{eq:dekker-condition} have to be satisfied.

We conclude from these considerations the remarkable fact that for the single-mode Gaussian map defined by the master \eqref{eq:gaussian_master}, mere positivity of the dynamics implies complete positivity.

\subsection{Gaussian map representation}
We can build upon this observation by employing the thorough analysis of general positive Gaussian maps presented in \cite{DePalma2015}, on which we rely upon repeatedly for the remainder of this subsection. 
Note, however, that in their work a convention for the covariance matrix is used that differs from ours by a factor of $2$.

We will rephrase the semigroup dynamics of \eqref{eq:gaussian_master} in terms of Gaussian maps. For that we need to investigate the dynamics of the variances \eqref{eq:covariance} induced by that master equation more closely. As the dynamics amounts to a phase space Ornstein-Uhlenbeck process, the corresponding evolution equation can be written as the differential Lyapunov equation \cite{Sandulescu_1986,SarkkaSolin2019}
\begin{equation}\label{eq:covdynamics}
    \dot\sigma = M\sigma + \sigma M^T + 2 D,
\end{equation}
with 
\begin{equation}\label{eq:covdynamicsmatrices}
   M = 
    \begin{pmatrix}
        \eta & 1\\
        -\Omega^2 & -\eta - 2\gamma
    \end{pmatrix}.
\end{equation}
Its solution for all times $t\ge 0$ is easily obtained:
\begin{equation}\label{eq:covdynamicsmap}
    \sigma(t) = X\sigma(0)X^T + Y,
\end{equation}
where
\begin{equation}
    X=X(t) = \e^{Mt}\;\;\mbox{and}\;\; Y=Y(t)=\int_0^tds\; \e^{M(t-s)}(2D)\e^{M^T(t-s)}.
\end{equation}

The covariance map \eqref{eq:covdynamicsmap} is of the form of the most general Gaussian quantum map \cite{Heinosaari2010, DePalma2015, Holevo2001, SerIllSie2003}. Conditions on the matrices $X$ and $Y$ that imply positivity or complete positivity are known and are thoroughly discussed in~\cite{DePalma2015}.
We note in passing that for our model of \eqref{eq:gaussian_master}, $X(t)$ and $Y(t)$ can be expressed explicitly in a simple form, from which all relevant properties can be deduced -- see Appendix \ref{app:covariance_propagator}. 

Let us first consider the issue of (mere) positivity for master \eqref{eq:gaussian_master}. We can repeat the argument by Dekker and Valsakumar in our formalism using the evolution \eqref{eq:covdynamics}: assume the dynamics is \emph{not} positive. Then, there is a time $t_0\ge 0$ at which some state loses positivity, i.e.\ $\det\sigma(t)<1/4$ for $t>t_0$ (yet close to $t_0$). Due to continuity, we must have $\det\sigma(t_0)=1/4$ and $\frac{d}{dt}\left.\det\sigma(t)\right|_{t=t_0}<0.$ 
Using simple steps as expanded in Appendix \ref{app:determinant}, 

we conclude from positivity violation at some time $t_0$ that
\begin{equation}\label{eq:dekker_inequality}
    0> \frac{\mathrm{d}}{\mathrm{d} t}\left.\det\sigma(t)\right|_{t=t_0} \ge
    -\gamma+2\sqrt{\det D}.
\end{equation}
We have thus proven that if conditions \eqref{eq:dekker-condition} are met, the dynamics is certainly positive.
We now use a result of \cite{DePalma2015} which states that such a Gaussian positive map in the form of \eqref{eq:covdynamicsmap} is also completely positive, if $\det X<1$. Here, indeed,
\begin{equation}
    \det X = \det (\e^{Mt}) = e^{\mathrm{tr}M t} = \e^{-2\gamma t}<1
\end{equation}
for all times $t>0$. In this way we understand more clearly that for a single mode, any merely positive Gaussian semigroup, as given in its most general form by the master equation \eqref{eq:gaussian_master} with a damping rate $\gamma>0$, is automatically completely positive. 

Later in Sec.~\ref{sec:qubit-semigroup} we will study an example of a qubit semigroup dynamics and find a very different behaviour, employing, essentially, similar tools adopted to the qubit case. There, the distinction between positive and completely positive dynamics turns out to be very relevant as we find positive qubit semigroups that are not completely positive.

\section{Quantum Brownian motion model and master equation}\label{sec:qbm_model}

There are only a few exact master equations known for genuine open system dynamics models. One of the most studied cases is the Ullersma model of a harmonic oscillator system, coupled bilinearly to an environment of harmonic oscillators \cite{Ullersma1966a, Ullersma1966b}. In the quantum domain, this model is also referred to as a \emph{quantum Brownian motion model} whose total Hamiltonian reads $H_{\rm tot}= H_{\rm sys}+ H_{\rm env}+ H_{\rm int}$ with
\begin{eqnarray}\label{eq:qbm_model}
H_{\rm sys} &=& \frac{p^2}{2} + \frac{1}{2}\Omega^2 q^2\,,\nonumber\\
H_{\rm env} &=& \sum_{i}\left(\frac{p_i^2}{2} +
\frac{1}{2}\omega_i^2 q_i^2\right),\\
H_{\rm int} &=& q\otimes B = q \otimes \sum_{i} g_i q_i\,. \nonumber
\end{eqnarray}
We are interested in the reduced dynamics,
\begin{equation}\label{eq:qbm_dynamicalmap}
    \rho(t) = \mbox{tr}_{\mathrm{env}}[U(t)(\rho(0)\otimes\rho_{\mathrm{env}})U^\dagger(t)]\equiv {\mathcal E}_{\mathrm{QBM}} (t)[\rho(0)],
\end{equation}
where $U(t) = \exp(-\i H_{\rm tot}t)$ and the initial state of the environment is assumed a thermal state, $\rho_{\mathrm{env}}=\exp(-\beta H_{\rm env})/Z$. Thus, ${\mathcal E}_{\mathrm{QBM}} (t)$ denotes the CPT dynamical map of the quantum Brownian motion model \eqref{eq:qbm_model}. As the total Hamiltonian is quadratic in all dynamical variables and the environmental initial state is Gaussian, ${\mathcal E}_{\mathrm{QBM}} (t)$ defines a Gaussian quantum channel for all $t$. 

The influence of the bath is encoded in the interaction picture bath coupling operator
\begin{equation}\label{eq:bcf_discrete}
    B(t) = \e^{\i H_{\rm env}t} \, B\;
\e^{-\i H_{\rm env}t},
\end{equation} with mean zero and the stationary correlation function
\begin{equation}\label{eq:bcf_continuous}
    \alpha(t-s) = \langle{B}(t){B}(s)\rangle= \mathrm{tr}[{B}(t){B}(s)\rho_{\mathrm{env}}(0)].
\end{equation}
In our work, $\alpha(t-s)$ is considered a continuous function of time that decays to zero strong enough for large $|t-s|$.

\subsection{Exact master equation}

It is well established that the dynamics of model \eqref{eq:qbm_model} is such that the reduced density 
operator $\rho(t)$ of \eqref{eq:qbm_dynamicalmap} satisfies an exact time local master equation  \cite{HaakeReibold1985,HuPazZhangPRD1992} for all times $t\ge 0$,
\begin{equation}
    \dot\rho(t) = \L_\mathrm{QBM}(t)[\rho(t)],
\end{equation}
with a \emph{time-dependent} generator that takes the form
\begin{eqnarray}\label{eq:qbm_masterosci}
\L_{\mathrm{QBM}}(t)[\rho] & = & -\i [H'(t),\rho]
- \i\Lambda(t) [q,\{p,\rho\}] \\ \nonumber
& & + 2d_{qp}(t) [q,[p,\rho]]
- d_{pp}(t) [q,[q,\rho]].
\end{eqnarray}
Here, $H'(t)=H_{\mathrm{sys}} + \frac{1}{2}\Delta(t) q^2$ is a Hamiltonian consisting of the original quadratic system Hamiltonian with a bath-induced shift $\Delta(t)$ of the oscillator frequency. Similar to the semigroup generator \eqref{eq:gaussian_master}, $\Lambda(t)$ is a (now time-dependent) damping rate, and $d_{qp}(t), d_{pp}(t)$ are (time-dependent) diffusion constants. These time-dependent functions are tricky to obtain, but can in principle be determined from evolution equations involving the parameters of the oscillator, and the bath correlation function $\alpha(t-s)$ from \eqref{eq:bcf_continuous}, see Ref.~\cite{HaakeReibold1985,HuPazZhangPRD1992,StrunzYu2004,Andras2023}. The dynamical map of \eqref{eq:qbm_dynamicalmap} thus takes the form of a time-ordered exponential of the integrated generator,
\begin{equation}
    {\mathcal E}_{\mathrm{QBM}}(t)=\mathcal{T}\exp\left\{\int_0^t ds\;\; \L_{\mathrm{QBM}}(s)\right\}.
\end{equation}

For our purpose it is important to note that whenever the bath correlation function $\alpha(t-s)$ decays strongly enough as $|t-s|$ grows, after a transient period $0\leq t\leq \tau$, all these time-dependent functions approach constant asymptotic values. The upper bound $\tau$ of that transient dynamics is typically of the order of the bath correlation time and thus much shorter than relevant system (relaxation) time scales.
Thus, we choose $\tau$ such that for $t > \tau$, we can replace the coefficients by constants,
\begin{eqnarray}
  \Delta(t)   & \longrightarrow & \tilde\Delta,  \\ \nonumber
   \Lambda(t)       & \longrightarrow & \tilde\Lambda, \\ \nonumber
        d_{qp}(t)       & \longrightarrow & \tilde d_{qp},  \\ \nonumber
   d_{pp}(t)  & \longrightarrow & \tilde d_{pp}.
\end{eqnarray}
Explicit expressions for these asymptotic values are known for specific bath correlation functions \cite{Andras2023} or for perturbative approaches to these\linebreak coefficients~\cite{Andras2023,StrunzHaake2003}. Thus, also the generator becomes time independent ${\L}_{\mathrm{QBM}}(t) \longrightarrow \tilde{\L}_{\mathrm{QBM}}$ for $t>\tau$ and takes the general Gaussian semigroup form \eqref{eq:gaussian_master} with $H=\tilde H'$, $\gamma = \tilde\Lambda$, $D_{qp} = \tilde d_{qp}, D_{pp} =  \tilde d_{pp}$, and, most notably, a zero $D_{qq}=0$.

We stress that according to the (complete) positivity condition \eqref{eq:dekker-condition}, $D_{qq}=0$ implies that the asymptotic generator of the dynamics does not belong to the GKSL class and thus, does not preserve positivity for all initial states. This holds true despite the fact that the semigroup master equation is derived from a unitary model of a joint system and environment dynamics.

\subsection{(Completely) positive semigroup for a restricted set of initial conditions}
\tikzstyle{densely dashed}=[dash pattern=on 5pt off 2pt]
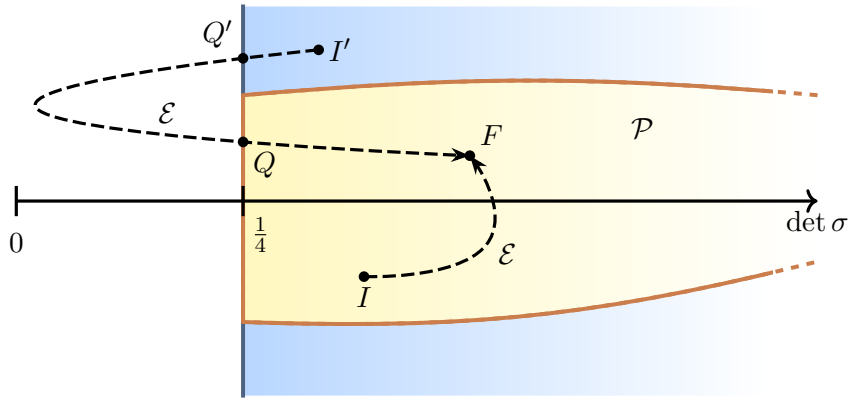
\begin{figure}
         \centering
        \begin{tikzpicture}[scale=2]
            \def\radius{1cm}
            \def\prad{1pt}
            \def\pure{1.5}
            \def\endfig{5}
            \coordinate (Origin) at (0,0);
            \coordinate (end) at (5.3,0);
            \coordinate (LU) at (\pure,-1.3);
            \coordinate (RO) at (\endfig,1.3);
            \coordinate (RU) at (\endfig,-1.3);
            \coordinate (LO) at (\pure,1.3);
            \coordinate (fix) at (3,0.3);
            \coordinate (outside) at (0.15,0.6);
            \coordinate (I) at (2,1);
            \coordinate (I2) at (2.3,-0.5);
            \coordinate  (inside) at (3.1,-0.3);
            \path[use as bounding box] (-0.5,-1.4) rectangle (6,1.4);
            
                \fill[line width=1.5pt, left color=NavyBlue!30!white, right color=white, draw=white] (LU) rectangle (RO);
                \draw[draw=NavyBlue!50!black, line width=1.5pt] (LU) -- (LO);
                \path[name path=rectangle] (LU) rectangle (RO);

                \def\plumcurve{(5.3,0.7)to[bend right=5] (\pure,0.7) to[bend right=0] (\pure,-0.8)    to[bend right=8] (5.3,-0.4)};
                \draw[Orange!80!black, left color=Goldenrod!30!white,right color=white,dashed,line width=1.5pt,line join=round]\plumcurve;
                \begin{scope}
                    \clip (-1,-2) rectangle (5,2);
                    \draw[Orange!80!black,line width=1.5pt,line join=round]\plumcurve;
                \end{scope}

                \node[above right] at (4,0.35) {$\mathcal{P}$};
                                
                \fill[black] (fix) circle [radius=\prad] node[above right] {$F$};
                \fill[black] (I) circle [radius=\prad] node[right] {$I^\prime$};
                \fill[black] (I2) circle [radius=\prad] node[below] {$I$};
                
                \draw[->, line width=1.2pt] (Origin)-- (end);
                    \draw[line width=1.2pt] (0,0.1) -- (0,-0.1) node[below=2pt] {$0$};
                    \draw[line width=1.2pt] (\pure,0.1) -- (\pure,-0.1) node[below right=-4pt and -1pt] {$\frac{1}{4}$};
                    \node[below] at (end) {$\det \sigma$};
                
                \path[name path=curve,draw=black,line width=1.2pt,arrows = {-Stealth[length=7pt]}, densely dashed] 
                    plot[smooth,        
                        tension=1] coordinates {(I) (outside) (fix)}; 
                \path[draw=black,line width=1.2pt,arrows = {-Stealth[length=7pt]}, densely dashed] 
                    plot[smooth,        
                        tension=1] coordinates {(I2) (inside) (fix)}; 
                \path[name intersections={of=curve and rectangle, by={L1,L2}}];
                        \fill[black] (L1) circle [radius=\prad];    
                        \node[above left] at (L1) {$Q^\prime$};
                        \fill[black] (L2) circle [radius=\prad];    
                        \node[below right] at (L2) {$Q$};

                \node[] at (1,0.58) {$\mathcal{E}$};
                \node[] at (3.25,-0.35) {$\mathcal{E}$};
        \end{tikzpicture}
    \caption{Schematic depiction of the space of physically admissible quantum states and mappings between them for the asymptotic master equation in \eqref{eq:qbm_masterosci_asymp}. Every valid quantum state with covariance matrix $\sigma$ has to satisfy the uncertainty relation, requiring  $\det \sigma \geq 1/4$. Under the evolution of a map $\mathcal{E}$ which features a single fixed point $F$, an initial condition can either remain positive for all times (example: $I$) or there is a time interval for which the map is not positive (example: $I'$). The point of the state space for the trajectory of the initial condition $I'$ to leave this space is indicated by $Q'$, it re-enters the state space at $Q$. Note that $Q$ itself as an initial condition would lead to positive dynamics as well while this is not true for $Q'$. The set of all initial conditions which yield positive dynamics for all times is indicated by the yellow region $\mathcal{P}$, the positivity domain. The color gradient depicts the mixedness of the state; intense colors (on the left) represent (nearly) pure states while white (on the right) represents highly mixed states.} 
    \label{fig:state_space_gaussian}
\end{figure}
After the transient period $0\le t \le \tau$, once the generator becomes time independent for $t>\tau$, the dynamics can be described by the asymptotic generator
\begin{eqnarray}\label{eq:qbm_masterosci_asymp}
\tilde\L_{\mathrm{QBM}}[\rho] & = & -\i [\tilde H',\rho]
- \i\tilde\Lambda [q,\{p,\rho\}] \\ \nonumber
& & + 2\tilde d_{qp} [q,[p,\rho]]
- \tilde d_{pp} [q,[q,\rho]].
\end{eqnarray}
Here, $\tilde H'=\frac{p^2}{2} + \frac{1}{2}(\Omega^2 + \tilde \Delta) q^2$. Again, we stress that this physically meaningful asymptotic generator of the dynamics is not of the desired (completely) positive GKSL class, and there is no need to be of that class: The generator is not supposed to act on \emph{any} initial system state $\rho(0)\in \states$, but only on the subset of  states $\rho(\tau)$ that have already undergone some dynamics during $0\le t\le \tau$.
In fact, for $t>\tau$, we can split the exact dynamics into an initial transient dynamics (the initial ``slip'' \cite{HaakeLewensteinPRA1983,StrunzDiosi1999} or ``jolt'' \cite{HuPazZhangPRD1992,FlemingHuRouraAoP2011,FlemingRouraHuPRE2011}), followed by semigroup dynamics with generator $\tilde\L_{\mathrm{QBM}}$. Thus, we can write (see also~\cite{StrachanPurkayasthaClarkJChemPhys2024})
\begin{equation}
    {\mathcal E}_{\mathrm{QBM}}(t)=\e^{\tilde\L_{\mathrm{QBM}}\cdot(t-\tau)}\circ \mathcal{T}\e^{\int_0^\tau ds\;\; \L_{\mathrm{QBM}}(s)}\;\;\mbox{for}\;\; t>\tau.
\end{equation}
We conclude from these considerations that the non-GKSL semigroup generator \eqref{eq:qbm_masterosci_asymp} describes asymptotic quantum dynamics of the quantum Brownian motion model. For that to hold true, we have to restrict the set of admissible initial states $\tilde\rho(0)\in {\mathcal P}_{\tilde\L_{\mathrm{QBM}}}$
to the image of the set of all states under the ``initial slip''-map,
\begin{equation}
    {\mathcal P}_{\tilde\L_{\mathrm{QBM}}} = \mathcal{T}\e^{\int_0^\tau ds\;\; \L_{\mathrm{QBM}}(s)} [\states],
\end{equation}
where $\states$ is the set of all (trace-class) initial states of the 1D-continuous variable system. By construction, all states from the restricted set $\tilde\rho(0)\in {\mathcal P}_{\tilde\L_{\mathrm{QBM}}}$ remain positive under semigroup dynamics with the non-GKSL master equation \eqref{eq:qbm_masterosci_asymp}. We call ${\mathcal P}_{\tilde\L_{\mathrm{QBM}}}$ the \emph{positivity domain} of the master \eqref{eq:qbm_masterosci_asymp}~\cite{JordanShajiSudarshanPRA2004}. This set is depicted by the yellow region in Fig.~\ref{fig:state_space_gaussian}. All initial states not in this positivity domain will evolve to this domain under the evolution of the exact (time-dependent) master \eqref{eq:qbm_masterosci} along a physical trajectory. However, using such a state $I'$ in Fig.~\ref{fig:state_space_gaussian} as the initial state for the semigroup \eqref{eq:qbm_masterosci_asymp} will result in dynamics violating positivity for a certain time interval.
We thus rephrase our characterization of the dynamics generated by the non-positive generator \eqref{eq:qbm_masterosci_asymp}: it is a positive semigroup generator for a \emph{restricted set of initial conditions} and thus, we call the semigroup dynamics \emph{restricted-positive}:
\begin{equation}\label{eq:rest_p}
\e^{\tilde\L_{\mathrm{QBM}} t}[\tilde\rho] \succeq 0 \;\;\mbox{for all}\;\; t\ge 0\;\;\mbox{and}\;\;\tilde\rho \in  {\mathcal P}_{\tilde\L_{\mathrm{QBM}}}.
\end{equation}

We can extend this idea to the notion of \emph{restricted completely positive} dynamics, enlarging our system Hilbert space by whatever auxiliary Hilbert space ${\mathcal H}_{\mathrm{A}}$ and choose a possibly entangled system-auxiliary initial state $\rho_{\mathrm{SA}}\in \states_{\mathrm{SA}}$. While the exact dynamical map
$ {\mathcal E}_{\mathrm{QBM}}(t)$ is completely positive, i.e.\
\begin{equation}\label{eq:full_cpt}
\left({\mathcal E}_{\mathrm{QBM}}(t)\otimes \id_{\mathrm{A}}\right) [\rho_{\mathrm{SA}}] \succeq 0 \;\;\mbox{for all}\;\; \rho_{\mathrm{SA}} \in \states_{\mathrm{SA}},
\end{equation}
the asymptotic semigroup generated by the non-GKSL $\tilde\L_{\mathrm{QBM}}$ will be restricted completely positive. Introducing the \emph{complete positivity domain} $c\mathcal{P}$, this is understood in the sense that
\begin{equation}\label{eq:rest_cpt}
\left(\e^{\tilde\L_{\mathrm{QBM}} t}\otimes \id_{\mathrm{A}}\right) [\tilde\rho_{\mathrm{SA}}] \succeq 0 \;\;\mbox{for all}\;\; t\ge 0\;\;\mbox{and}\;\;\tilde\rho_{\mathrm{SA}} \in  c{\mathcal P}_{\tilde\L_{\mathrm{QBM}}},
\end{equation}
where the correlated and possibly entangled admissible initial system-auxiliary states $\tilde\rho_{\mathrm{SA}}$  for restricted completely positive dynamics have to be taken from the ``initial slip'' subset
\begin{equation}\label{eq:rest_cpt_domain}
    c{\mathcal P}_{\tilde\L_{\mathrm{QBM}}} = \left(\mathcal{T}\e^{\int_0^\tau ds\;\; \L_{\mathrm{QBM}}(s)}\otimes \id_{\mathrm{A}}\right) [\mathcal{S}_{\mathrm{SA}}],
\end{equation}
which is a strict subset of the set of all density operators $\mathcal{S}_{\mathrm{SA}}$ acting on ${\mathcal H}_{\mathrm{S}}\otimes{\mathcal H}_{\mathrm{A}}$.

\section{A family of non-GKSL Qubit semigroup dynamics}\label{sec:qubit-semigroup}

\subsection{Master equation of GKSL-type}

The quantum Brownian motion model discussed above describes a\linebreak continuous-variable system. In order to understand the phenomenon of physically meaningful semigroup dynamics which are not (completely) positive in a simpler setting, we aim at considering the smallest possible quantum system, a qubit with Hilbert space dimension $d=2$. Thus, we are interested in situations where we loosen the requirement that a qubit semigroup needs to be (completely) positive for all possible initial states: we now consider dynamics where only a certain subset of initial states remains (completely) positive under the dynamics. In fact, we will focus on simple positivity for most of this section and comment on complete positivity only at the very end.

This leads us to the notion of restricted positivity, now for qubit dynamics. As before, we say a map $\map: \states(\mathcal{H}_d) \to \states(\mathcal{H}_d)$ with $\states$ being the set of all physical qubit quantum states is \emph{restricted positive}, if there is a non-empty set of admissible quantum states $\mathcal{P} \subseteq \states$ which remain positive under the application of the map $\map$, meaning $\mathcal{P}=\{\rho| \rho\in \states,\ \map[\rho] \succeq 0\}$. For a family of quantum maps $\map_t$ with a parameter $t\geq0$ representing time, restricted positivity of $\map_t$ means that there is a set $\mathcal{P}\subseteq\states$ of initial states $\rho(0)$ for which $\rho(t) = \map_t[\rho(0)]$ remains positive for all times $t\geq0$.
The largest possible set $\mathcal{P}$ is then termed the \emph{positivity domain}. 
Clearly, for a contractive semigroup $\map_t=\exp(\L t)$ with an asymptotic stationary state (a fixed point), the positivity domain is never empty -- this is the case we are going to study here.

Notice that if $\mathcal{P}$ is identical to the set of all quantum states, the dynamics is positive. For the sake of clarity, we refer to dynamics in which $\mathcal{P}$ encompasses all states as positive, and to those in which $ \mathcal{P}$ is a proper subset of all states as restricted positive.

We will now discuss a family of semigroup dynamics which shows completely positive, (merely) positive and restricted positive dynamics depending on the choice of certain parameters of the generator.
Using the notation $\mathcal{G}_{i}^{j}[\rho]=\sigma_i \rho \sigma_j -\frac{1}{2}\{\sigma_j \sigma_i,\rho\}$ we propose the generator 
\begin{equation}\label{eq:CigarGenerator}
    \mathcal{G}[\rho]=\dot{\rho}=\left(1-\frac{\alpha}{2}\right) \mathcal{G}_{x}^{x}[\rho] -\frac{\alpha}{2}\left(\mathcal{G}_{y}^{y}[\rho]+\mathcal{G}_{z}^{z}[\rho]\right) +\frac{\i \beta}{2} \left(\mathcal{G}_{y}^{x}[\rho]-\mathcal{G}_{x}^{y}[\rho]\right),
\end{equation} 
where $\alpha$ and $\beta$ are real-valued parameters.
Since this generator is time-independent, a simple integration yields a one-parameter semigroup of maps $\{\mathcal{E}_t\}$
\begin{equation}
    \mathcal{E}_t[\rho]=\e^{\mathcal{G }t}[\rho].
\end{equation}

We can equally express this generator in a more familiar way, using the canonical form of a master equation \cite{HalCreLiAnd2014}, yielding
\begin{equation}\label{eq:CigarGenerator2}
    \mathcal{G}[\rho]=\dot{\rho}=\sum_{i=1}^{d^2-1} \gamma_i \left(J_i \rho J_i^\dagger -\frac12 \{J_i^\dagger J_i, \rho\}\right).
\end{equation}
Here the $J_i$ take the role of jump-type operators and the corresponding $\gamma_i$ represent the decay rates. 
For a generator to be of the GKSL form from \eqref{eq:gksl_master} all decay rates $\gamma_i$ have to be positive and constant and only then it is possible to define proper jump operators $L_i$ as $L_i=\sqrt{\gamma_i} J_i$ in \eqref{eq:gksl_master}. If \eqref{eq:CigarGenerator} is not of GKSL-form, we say this equation is of GKSL-\emph{type}~\cite{Lin1976b, GorKosSud1976}, and at least one of the $\gamma_i$ is negative.

In correspondence with \eqref{eq:CigarGenerator} the decay rates become $\gamma_1=\frac{\alpha}{2}$ and $\gamma_{2,3}=\frac{1}{2}\left(1\pm \sqrt{(\alpha-1)^2+\beta^2}\right)$
and the jump operators are given by: 

\begin{subequations}
   \begin{align}
    J_1&=\frac{1}{\sqrt2}\sigma_z, \label{eq:LindbladtypeOperator1}\\ 
    J_2&=\frac{ \beta}{2\sqrt{c} \sqrt{c+\alpha-1}}\sigma_x+\frac{ \i \sqrt{c+\alpha-1}}{2\sqrt{c}}\sigma_y,\label{eq:LindbladtypeOperator3}\\
    J_3&= \frac{ \beta}{2\sqrt{c} \sqrt{c-\alpha+1}}\sigma_x -\frac{\i \sqrt{c-\alpha+1}}{2\sqrt{c}}\sigma_y,\label{eq:LindbladtypeOperator2} 
\end{align}
\end{subequations}
where we use the abbreviation $c=\sqrt{(\alpha-1)^2+\beta^2}$ and the $\sigma_i$ are the Pauli matrices with $\sigma_0=\id$.

Given a suitable choice of the real parameters $\alpha$ and $\beta$, it turns out that the dynamics is either i) completely positive, ii) positive or iii) truly restricted positive. 
Only in case i) the generator \eqref{eq:CigarGenerator2} yields a GKSL master equation, in the other two cases, the dynamics are only of GKSL-type. We are especially interested in the latter case, but will provide a short discussion of all cases i) - iii) for completeness.

\subsection{Bloch representation of the dynamics}
As we concern ourselves with qubit dynamics, a useful geometric interpretation of the dynamics can be obtained using the Bloch representation.
Similar to the Bloch representation of a density matrix $\rho$ in terms of the Bloch vector $\vec{r}$ given by
\begin{equation}
    \vec{r}=\big(\tr{(\rho\sigma_x)},\tr{(\rho\sigma_y)},\tr{(\rho\sigma_z)}\big)^T,
\end{equation}
a qubit map can also be written in the Pauli basis. For the generator in \eqref{eq:CigarGenerator},
the expansion in Pauli matrices -- reintroducing $\sigma_0=\id$ -- yields the $4\times4$ matrix
\begin{equation}
    \Gamma_{ij}=\frac12 \tr{\sigma_i \mathcal{G}[\sigma_j]},
\end{equation}
or written as a  matrix
\begin{equation}
    \Gamma=\begin{pmatrix}
        0&\vec{0}^T\\
        \vec{c}& G
    \end{pmatrix}.
\end{equation}
Here, $G$ is a $3\times3$ real matrix and $\vec{c}$ is a three component vector.
The action of a master equation on the Bloch vector thus takes the form
\begin{equation}\label{eq:Blochgenerator}
    \dot{\vec{r}}(t)= G \vec{r}(t) + \vec{c}.
\end{equation}
 Integrating this ordinary differential equation gives the map $\Lambda_t$ in Bloch representation \cite{bethruskaiAnalysisCompletelypositiveTracepreserving2002}
\begin{equation}\label{eq:Blochmap}
    \vec{r}^\prime=\vec{r}(t)= \Lambda_t[\vec{r}(t_0)]=T_t \vec{r}(t_0) + \vec{\tau}_t,
\end{equation}
where $T_t=\e^{Gt}$ is a $3\times3$ real matrix and $\vec{\tau}_t = (T_t-\id)G^{-1}\vec c$ is a three component vector that describes the shift of the Bloch ball center under the dynamics $\Lambda_t[\vec{0}]=\vec{r}_0(t)= \vec{\tau}_t$. 
The deformation matrix $V=TT^T $ provides the evolution of the shape of the Bloch sphere. Its eigenvectors determine the orientation of the axes of the ellipsoids and the eigenvalues are the squares of the ellipsoid semi axes~\cite{milneQuantumSteeringEllipsoids2014}.

For our particular example, we obtain from the rates $\gamma_i$ and the jump-type operators 
\eqref{eq:LindbladtypeOperator1} -- \eqref{eq:LindbladtypeOperator2} the objects
\begin{equation}
G=\begin{pmatrix}
    -\alpha &0 &0\\0&-1&0\\0&0&-1
\end{pmatrix}, \qquad \vec{c}=\begin{pmatrix}
    0\\0\\\beta
\end{pmatrix}.  
\end{equation}
This leads to a quantum map in Bloch representation with
\begin{equation}
    T_t=\begin{pmatrix}
   \e^{-\alpha t} &0 &0\\0&\e^{-t}&0\\0&0&\e^{- t}
\end{pmatrix}, \qquad \vec{\tau}_t=\begin{pmatrix} 0\\0\\\beta(1-\e^{-t})
\end{pmatrix}.
\end{equation}
The eigenvalues of the deformation matrix $V$ are $\{\e^{-2t},\e^{-2t}, \e^{-2\alpha t}\}$. For all three semi-axes of the time-evolving ellipsoid to be reduced in length, $\alpha$ must be positive and thus the dynamics are contractive. Furthermore, this dynamics features a fixed point given by $\lim_{t\rightarrow\infty} \vec{\tau}_t=(0,0,\beta)^T$. To ensure that this fixed point is a valid quantum state we assert that $-1 \leq \beta \leq 1$. For $\alpha>0$ and such $\beta$, the rates $\gamma_1$ and $\gamma_2$ are always positive. It is the sign and value of $\gamma_3$, as depending on the two parameters $\alpha$ and $\beta$, that determines the type of semigroup.

\subsection{Different types of positivity}
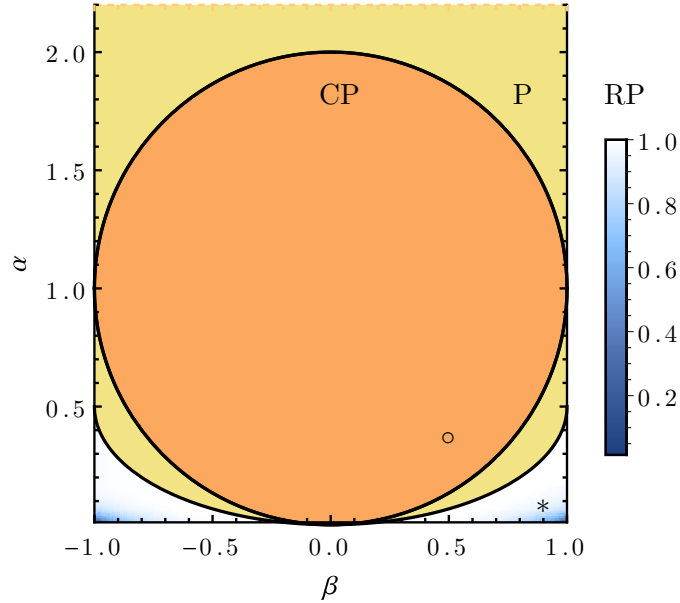
\begin{figure}[!htb]
    \centering
        \begin{tikzpicture}
    \node[anchor=south west,inner sep=0] (image) at (0,0) {\includegraphics[width=0.7\linewidth]{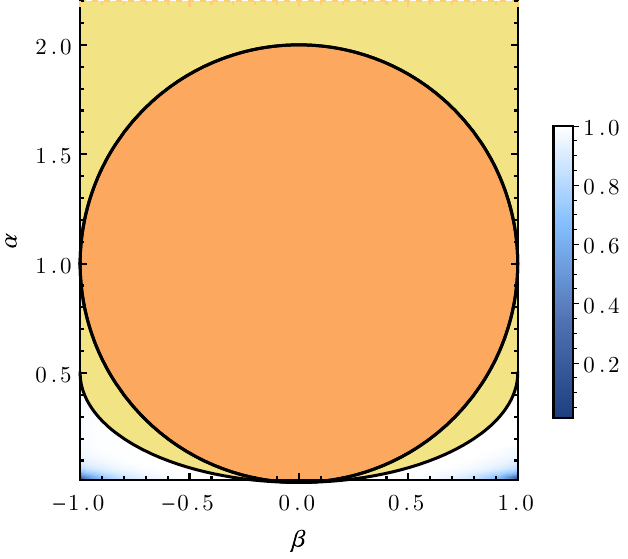}};
    \begin{scope}[x={(image.south east)},y={(image.north west)}]
        \node at (0.49,0.85) {CP};
        \node at (0.76,0.85) {P};
        \node at (0.91,0.85) {RP};
        \node at (0.65,0.28) {{\hyperref[fig:Depolarizingchannel]{$\circ$}}};
        \node at (0.79,0.167) {{\hyperref[fig:Zigarrendynamik]{$\ast$}}};
    \end{scope}
\end{tikzpicture}
    \caption{The figure shows the $(\beta,\alpha)$-parameter plane. All dynamics with the parameters $(\beta,\alpha)$ from the orange circle are completely positive dynamics. The yellow region corresponds to parameters for which the dynamics are positive. Note that the parameter plane extends to all positive $\alpha$ and all parameter choices with $\alpha>2$ correspond to positive dynamics. The color gradient from white to blue represents the volume of the positivity domain $\mathcal{P}$ relative to volume of the Bloch ball for restricted positive dynamics. The parameter values marked by an open circle ($\circ$) and an asterisk ($\ast$) correspond to the dynamics depicted in Fig.~\ref{fig:Depolarizingchannel} and Fig.~\ref{fig:Zigarrendynamik}, respectively.
    }
    \label{fig:ParameterPlane}
\end{figure}

We will now investigate under which parameter choices the dynamics described by \eqref{eq:CigarGenerator} is completely positive, positive or only restricted positive, as depicted in Fig.~\ref{fig:ParameterPlane}.

\subsubsection{Case i: Completely Positive Dynamics}
A dynamics is completely positive if and only if all decoherence rates $\gamma_i$ of the corresponding master GKSL-type master \eqref{eq:CigarGenerator2} are positive \cite{GorKosSud1976}.

\begin{figure}[!htb]
    \centering
    \begin{subfigure}[b]{0.32\textwidth}
        \includegraphics[width=0.95\linewidth]{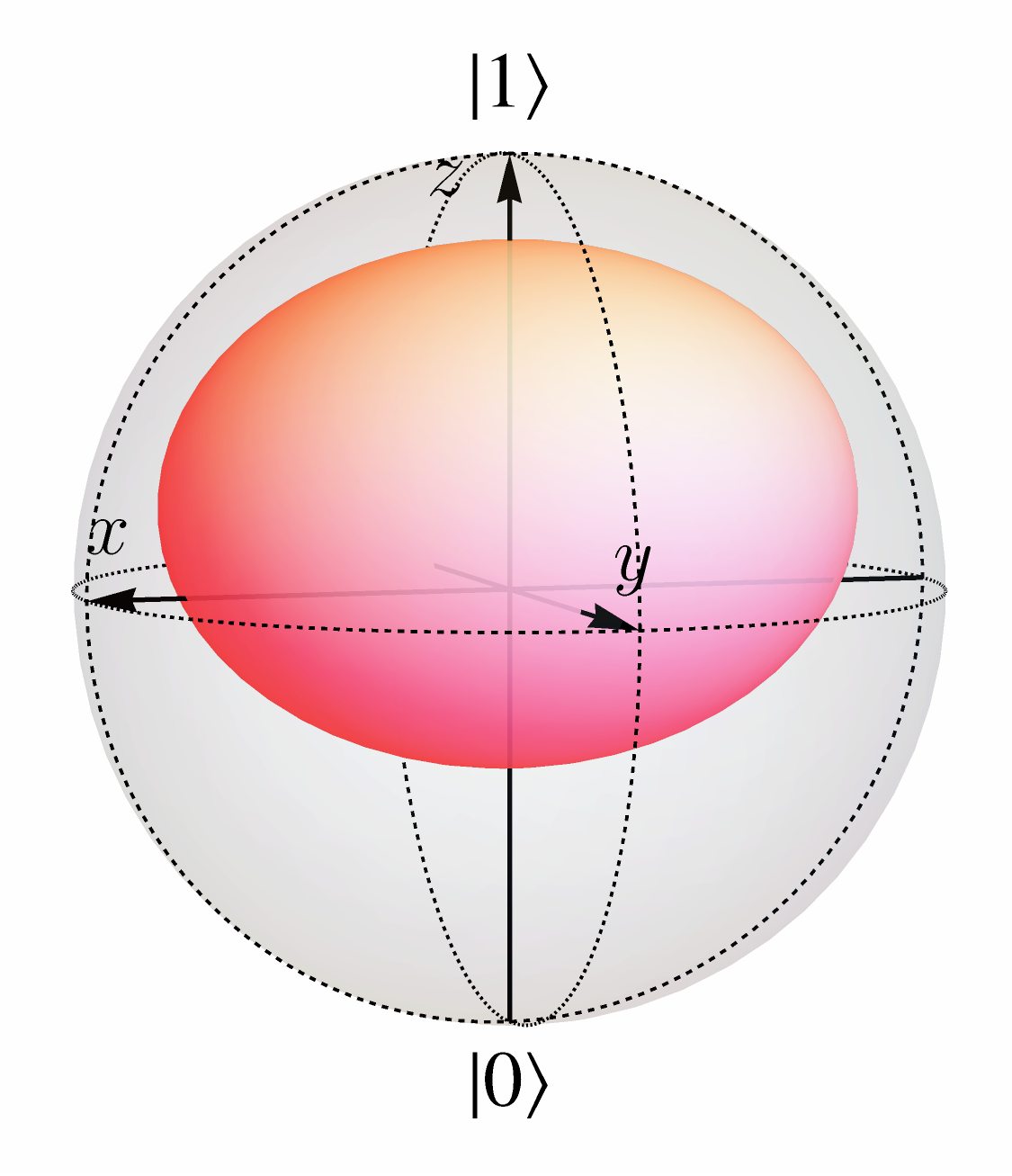}
        \caption{$t=\frac12$.}
    \end{subfigure}
    \begin{subfigure}[b]{0.32\textwidth}
        \includegraphics[width=0.95\linewidth]{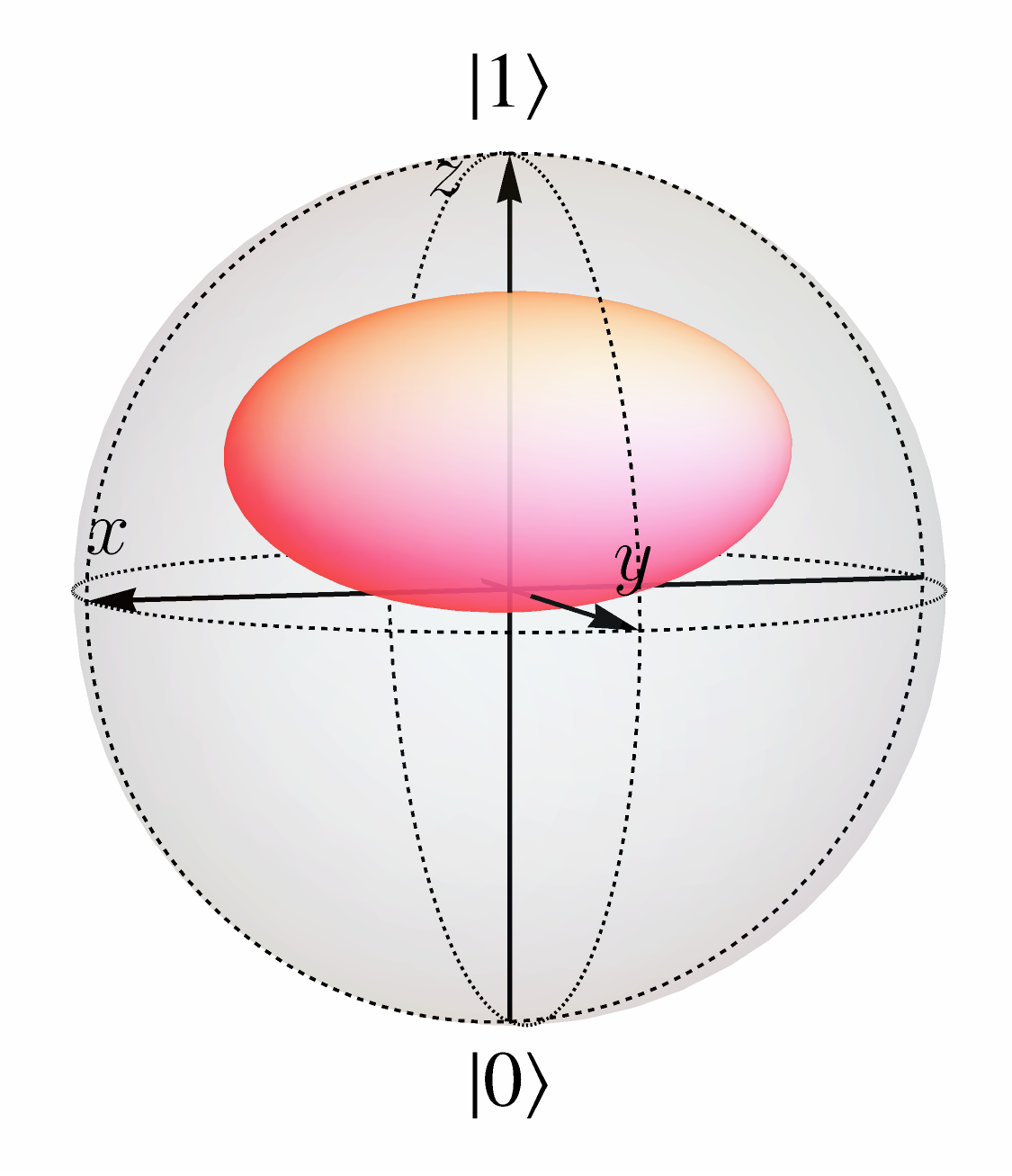}
        \caption{$t=1$.}
    \end{subfigure}
    \begin{subfigure}[b]{0.32\textwidth}
        \includegraphics[width=0.95\linewidth]{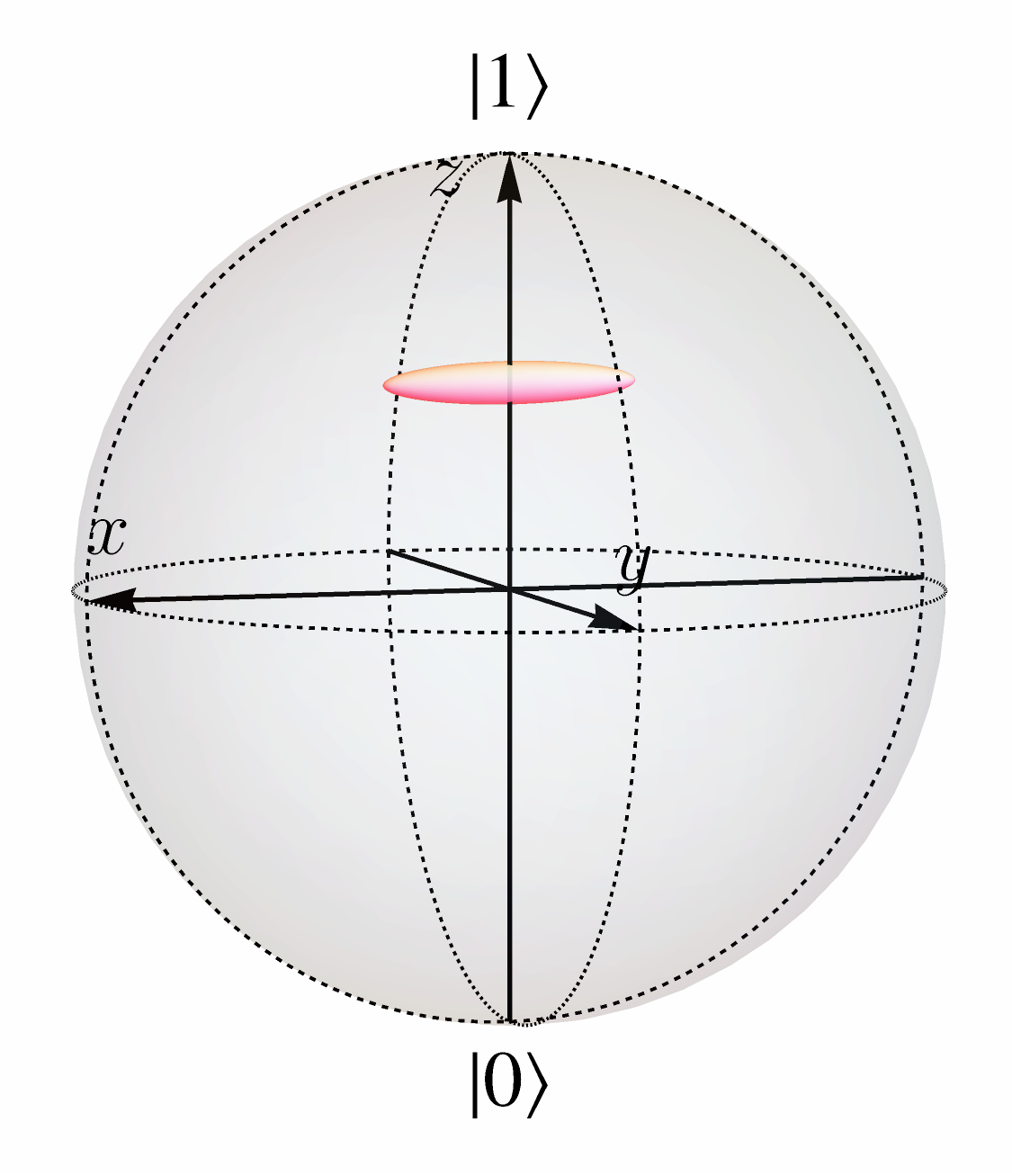}
        \caption{$t=3$.}
    \end{subfigure}
    \caption{
The figure illustrates the image of the time-evolved Bloch ball, $\Lambda_t \mathcal{B}$ at various points in time, with parameter choices $\alpha=\frac{4}{10}$ and $\beta=\frac{1}{2}$ (open circle ($\circ$) in Fig.~\ref{fig:ParameterPlane}). The $y$- and $z$-axes of the Bloch sphere contract at the same rate, while the $x$-axis contracts more slowly. The fixed point of the dynamics is $F=(0,0,\frac{1}{2})^T$. Obviously, the image of the Bloch sphere remains within the Bloch sphere for all times $t>0$.}
    \label{fig:Depolarizingchannel}
\end{figure}

For our dynamics this condition is satisfied if $\gamma_3 \ge 0$ which implies
\begin{equation}\label{eq:cpt_condition}
    (\alpha-1)^2+\beta^2 \le 1.
\end{equation}
Consequently, the parameters that result in a completely positive dynamics describe a unit circle in the parameter space $(\beta,\alpha)$ centered at $(0, 1)$, as depicted in Fig.~\ref{fig:ParameterPlane}. We note in passing that for the choice $\alpha=1$ and $\beta=0$ the generator describes an instance of the depolarizing channel  
\begin{equation}
    \mathcal{E}_\mathrm{D}[\rho]=(1-\lambda)\rho + \frac\lambda2 \id,
\end{equation}
with $\lambda=1-\e^{-t}$. However, more complex channels can also be realized, as shown in Fig.~\ref{fig:Depolarizingchannel} for the parameter choices $\alpha= 0.4$ and $\beta=0.5$, which are marked by an open circle ($\circ$) in Fig.~\ref{fig:ParameterPlane}.

\subsubsection{Case ii: Positive Dynamics}
Any completely positive dynamics is, by definition, also positive. The converse, however, does not hold and we will hence investigate positivity with a focus on those parameter choices yielding positive but not completely positive dynamics. In contrast to the Gaussian case considered in Section \ref{sec:qbm_master}, we will see here that for qubits positivity of a semigroup does not imply complete positivity.

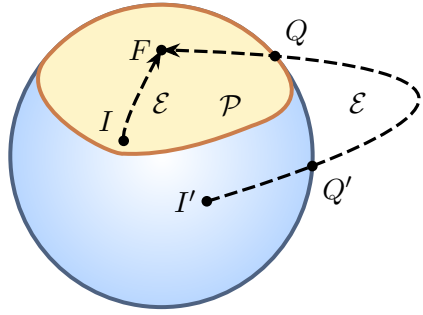
\begin{figure}[!htb]
    \centering
       \begin{tikzpicture}[scale=2]
            \def\radius{1cm}
            \def\prad{1pt}
            \coordinate (center) at (0,0);
            \coordinate (fix) at (0,0.7);
            \coordinate (I) at (0.3,-0.3);
            \coordinate (outside) at (1.7,0.4);

            \def\mycurve{ plot[smooth,tension=1] coordinates {(I) (outside) (fix)}};
            
            \draw[NavyBlue!50!black,outer color=NavyBlue!30!white, inner color=white, line width=1.5pt] (center) circle [radius=\radius];
            
            \path[name path=circle] (center) circle [radius=\radius];
            \path[name path=curve] \mycurve;

            \path[name intersections={of=curve and circle, by={L1,L2}}];

        \coordinate (A) at (140:\radius);
        \coordinate (B) at (0.6,0.2);
        \coordinate (C) at (-0.25,0.015);
        \coordinate (C2) at (-0.25,0.1);
        \coordinate (inside) at (-0.2,0.3);

                \begin{scope}
                    \path[
                        fill=Goldenrod!30!white,
                        draw=Orange!80!black,
                        line width=1.5pt
                    ]
                    (A)
                    arc (140:40:\radius)
                    .. controls (1.00,0.36) and (0.75,0.25)
                    .. (B) 
                    .. controls (0.35,0.10) and (0.05,0.00)
                    .. (C)
                    .. controls (-0.35,0.00) and (-1.00,0.35)
                    .. (A);
                \end{scope}

            \node[above left] at (B) {$\mathcal{P}$};
            \path[draw=black,line width=1.2pt,arrows=-{Stealth[length=7pt]},densely dashed] \mycurve;
            \fill[black] (I) circle [radius=\prad]node[left]  {$I^\prime$};
            
            \fill[black] (C2) circle [radius=\prad] node[above left]  {$I$};

            \path[name path=curvein,draw=black, line width=1.2pt,arrows = {-Stealth[length=7pt]}, densely dashed] 
                plot[smooth,        
                    tension=1] coordinates {(C2) (inside) (fix)};   

            \fill[black] (fix) circle [radius=\prad] node[left]  {$F$};
            \fill[black] (L1) circle [radius=\prad];    \node[below right] at (L1) {$Q^\prime$};
            \fill[black] (L2) circle [radius=\prad];    \node[above right] at (L2) {$Q$};

            \node[] at (1.3,0.35) {$\mathcal{E}$};
            \node[] at (0,0.35) {$\mathcal{E}$};
        \end{tikzpicture}
    \caption{Sketch of the situation of a non-positive qubit dynamics. The blue circle represents the Bloch ball and the yellow subset thereof the positivity domain $\mathcal{P}$. The trajectory of $I^\prime$ (and $Q^\prime$) leave the Bloch ball for a while, then eventually approaches the fixed point $F$. All initial conditions from within the positivity domain give rise to positive dynamics.}
    \label{fig:state_space_qubit}
\end{figure}

Considering Fig.~\ref{fig:state_space_qubit}, suppose there exists some initial condition $I^\prime\in \mathcal{B}$ with $\mathcal{B}$ representing the full qubit state space in terms of the Bloch ball, that does not preserve positivity. Thus, at some time $t_0 >0$, the trajectory of $I'$ leaves the Bloch ball: $\Lambda_{t}[I^\prime]\notin \mathcal{B}$ for $t>t_0$ (but close to $t_0$).
Since it is a continuous mapping, $t_0$ is the time where the trajectory of $I'$ intersects the Bloch sphere in some point $Q^\prime$. By the semigroup property, if the trajectory subsequently leaves the Bloch ball, then $Q^\prime$ must be mapped outside the Bloch ball after an infinitesimal time step. As we consider contractive dynamics with a single fixed point $F$, ultimately every state is mapped onto $F$. Hence, the trajectory of $I^\prime$ (and $Q^\prime$) must intersect the Bloch sphere at some later time in a point~$Q$. At this later intersection, the trajectory heads back into the Bloch ball, thereby giving rise to positive dynamics from that point onward. It is worth noting that the point $Q$ represents a pure state which gives rise to positive dynamics. This implies that the positivity domain $\mathcal{P}$ must always contain at least one pure state.

In summary, if an initial condition leads to non-positive dynamics, then there must exist a pure state that is mapped outside the Bloch sphere in an infinitesimal time. Therefore, we can distinguish between positive and non-positive dynamics by examining the direction of the velocity vector $\dot{\vec{r}}$ of all pure states $|\vec{r}|=1$. 
If the velocity vector points outward from the Bloch ball, then $\vec{r}\cdot\dot{\vec{r}}>0$ and the dynamics is not positive.

Plugging in the master equation \eqref{eq:Blochgenerator}, the condition for positive dynamics thus reads
\begin{equation}\label{eq:DirectionVelocityVector}
    \vec{r}\cdot\dot{\vec{r}}=\vec{r}^T G \vec{r}+\vec{r}\cdot\vec{c} \le 0\;\;\mbox{for all}\;\;|\vec r|=1.
\end{equation}
In terms of our parameters $\alpha$ and $\beta$, this implies that the dynamics is positive for all $\alpha$ above the curve
\begin{equation}\label{eq:BoundaryPRP}
    \alpha\ge \alpha(\beta)=\frac{1}{2}\left(1-\sqrt{1-\beta^2}\right),
\end{equation} 
which can be seen as bounding curves of the plum-colored positivity region at the bottom in Fig. \ref{fig:ParameterPlane}.
For more details on this boundary, see Appendix \ref{sec:appendixRP_P}.

Note again that in our example of qubit dynamics not every parameter choice leading to positive dynamics also leads to completely positive dynamics, see also Fig.~\ref{fig:ParameterPlane}. This is in contrast to what we have observed before in Sec.~\ref{sec:complete_positivity_condition} in the context of Gaussian dynamics, where generators that lead to positive dynamics are precisely those corresponding to completely positive dynamics (i.e.\ GKSL generators).

\subsubsection{Case iii: Restricted Positive Dynamics}

\begin{figure}[!htb]
    \centering
    \begin{subfigure}[b]{0.32\textwidth}
        \includegraphics[width=0.95\linewidth]{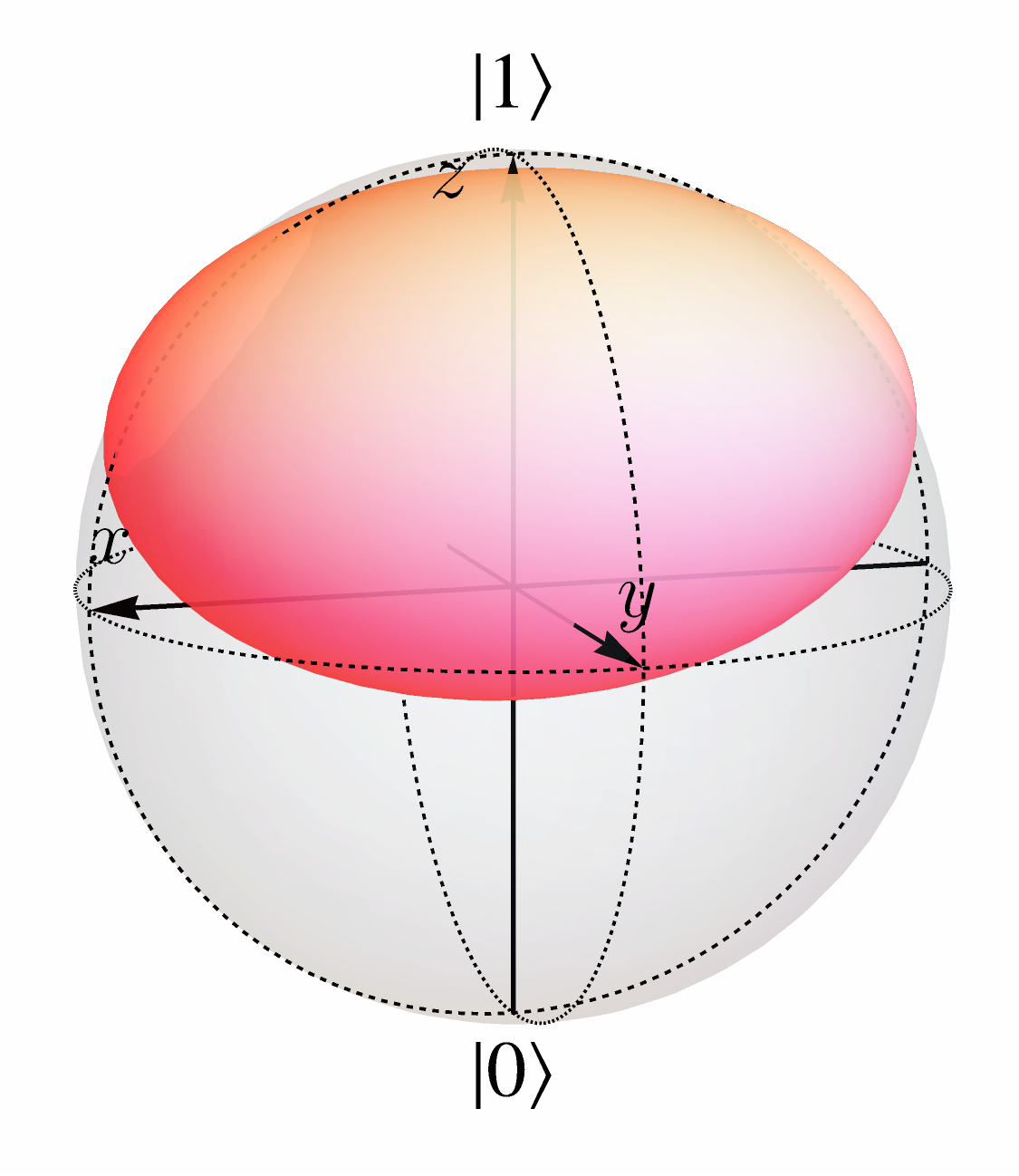}
        \caption{$t=\frac12$.}
    \end{subfigure}
    \begin{subfigure}[b]{0.32\textwidth}
        \includegraphics[width=0.95\linewidth]{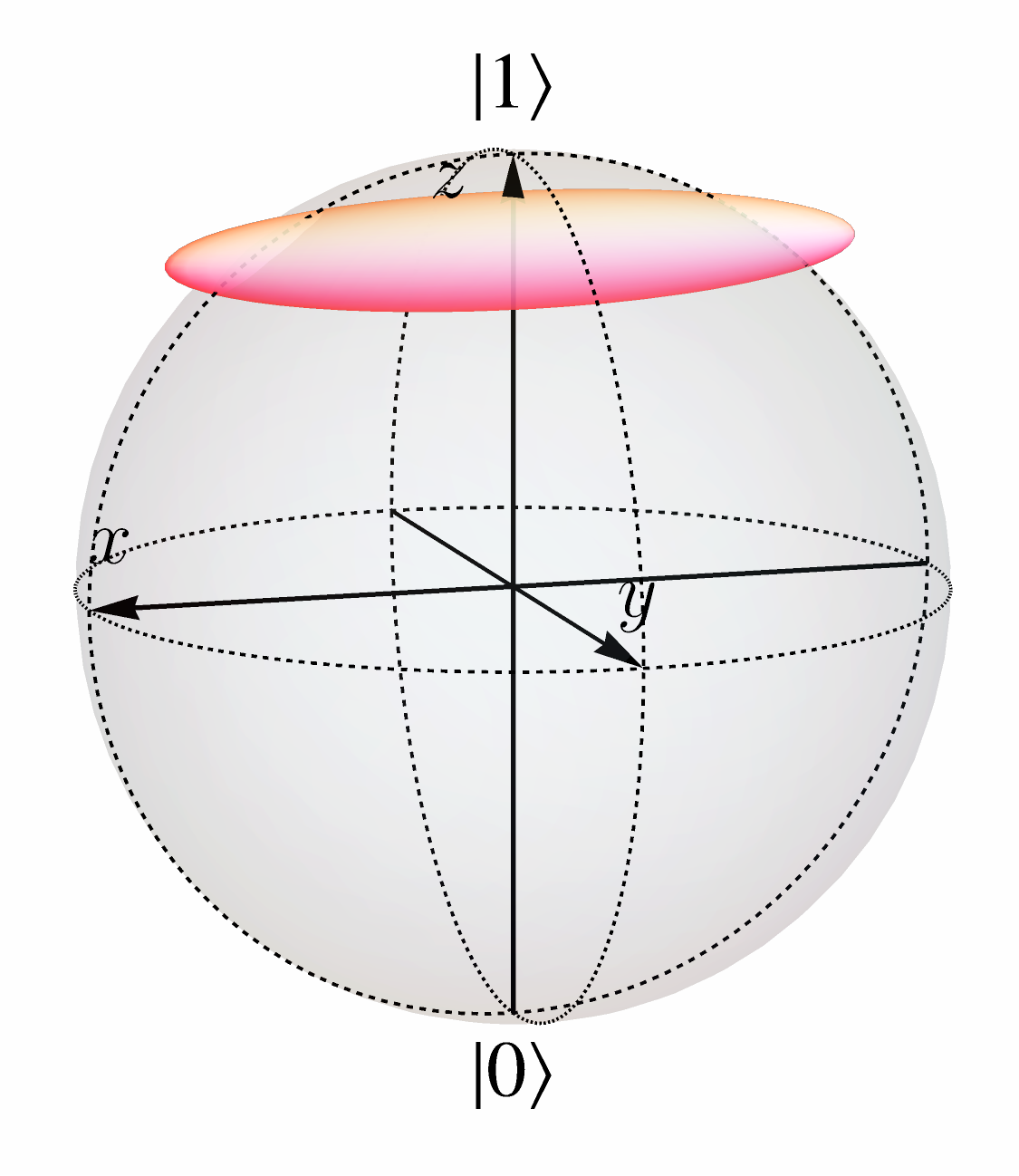}
        \caption{$t=2$.}
    \end{subfigure}
    \begin{subfigure}[b]{0.32\textwidth}
        \includegraphics[width=0.95\linewidth]{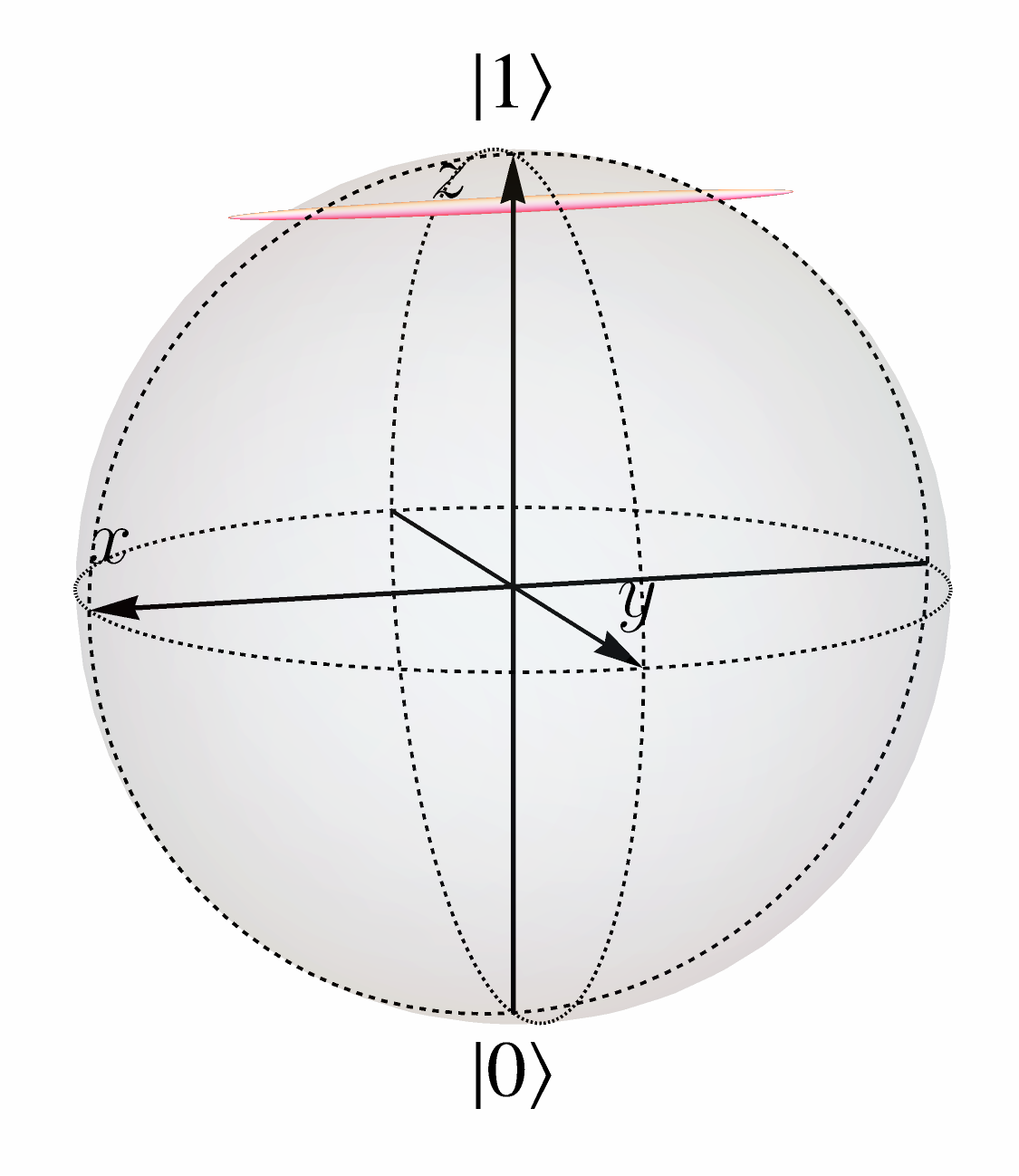}
        \caption{$t=4$.}
    \end{subfigure}
    \caption{
Shown here is an example of a restricted positive dynamics according to the generator from \eqref{eq:CigarGenerator} for the parameter values $\alpha=\frac{1}{10}$ and $\beta=\frac{9}{10}$, marked by an asterisk~($\ast$) in Fig.~\ref{fig:ParameterPlane}. The figure illustrates the image of the time-evolved Bloch ball, $\Lambda_t \mathcal{B}$, at three different times $t$. The $y$- and $z$-axes of the Bloch sphere contract at the same rate, while the $x$-axis contracts much more slowly. At the same time, the ellipsoid shifts toward the fixed point $F=(0,0,\frac{9}{10})^T$. Some initial states leave the Bloch sphere under the dynamics and therefore do not yield positive dynamics, others remain in the Bloch ball for all times.}
    \label{fig:Zigarrendynamik}
\end{figure}

We have found that all parameter choices $(\beta,\alpha)$ below the curve given by $\alpha(\beta)=\frac{1}{2}\left(1-\sqrt{1-\beta^2}\right)$ lead to dynamics with at least one initial condition that does not give rise to positive dynamics. An example for such dynamics is illustrated in Fig.~\ref{fig:Zigarrendynamik}; this corresponds to parameters indicated by an asterisk~($\ast$) in Fig.~\ref{fig:ParameterPlane}.

\begin{figure}[!htb]
    \centering
    \includegraphics[width=0.5\linewidth]{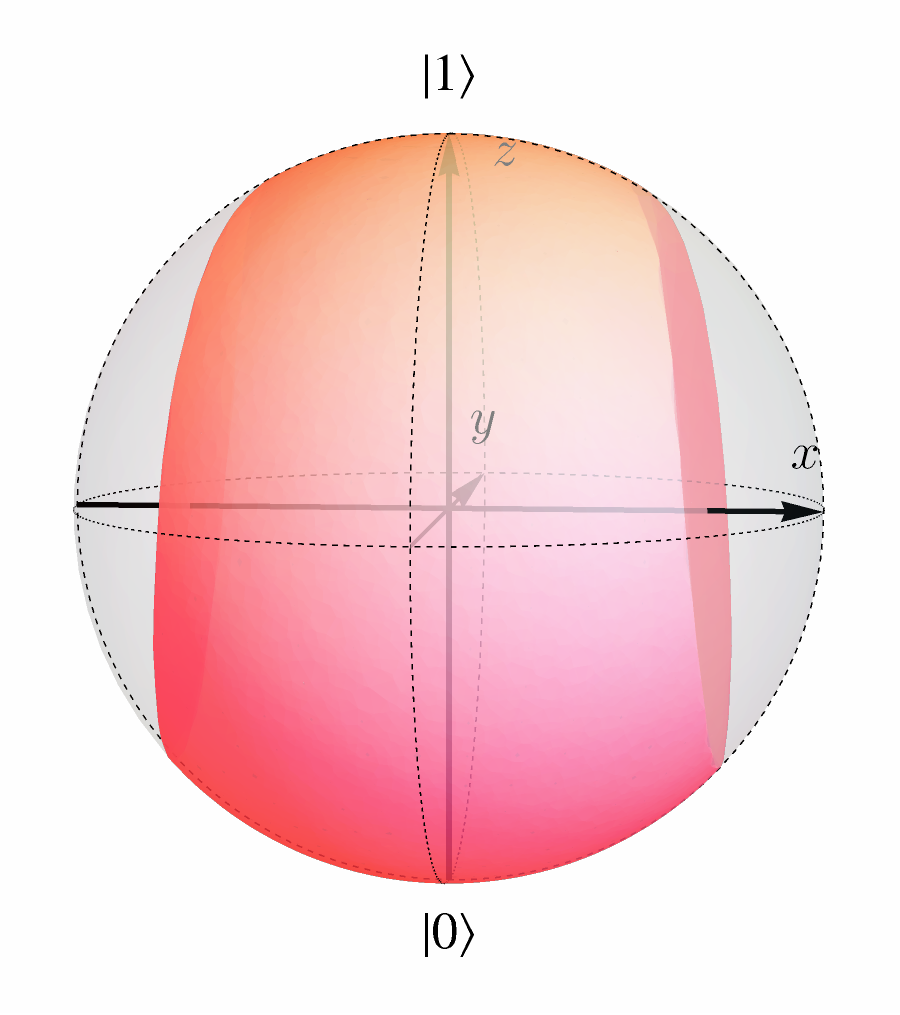}
    \caption{Depiction of the positivity domain for dynamics according to \eqref{eq:CigarGenerator} with parameter $\beta=\frac{9}{10}$ and $\alpha=\frac{1}{10}$. In Fig.~\ref{fig:ParameterPlane} this is indicated by an asterisk ($\ast$) in the area showing restricted positivity. This means that about $86.7\%$ of all possible initial states remain positive under the evolution of this map.}
    \label{fig:PositiveDomain}
\end{figure}

We have seen above that there always exists at least one pure initial state that gives rise to positive dynamics. However, for the model to have genuine physical significance, it is not sufficient for positive dynamics to occur only under a single isolated initial condition. There should rather be a set of initial conditions with positive measure that evolve with positive dynamics. 
To identify such sets of positive measure, we utilize the fact that the space of quantum states is convex.
For an analytical approach to check for the existence of such a set see Appendix \ref{sec:appendix_tetrahedron}. A numerical analysis yields Fig.~\ref{fig:PositiveDomain}, which depicts the positivity domain of the restricted positive dynamics obtained from \eqref{eq:CigarGenerator} for parameters $\alpha=0.1$ and $\beta=0.9$, marked by an asterisk ($\ast$) in Fig.~\ref{fig:ParameterPlane}. 
The size of the positivity domain depends on the parameters $\alpha$ and $\beta$. As the parameters approach the boundary of the positive dynamics region, the positivity domain grows. The gradient in the density plot from Fig.~\ref{fig:ParameterPlane} indicates the volume of the positivity domain relative to the volume of the Bloch ball.

Positivity is not sufficient to guarantee physical dynamics in the conventional sense. Since we consider open quantum systems, the system of interest may be correlated with auxiliary degrees of freedom. The dynamics must therefore remain positive not only for the system itself, but also when extended to an ancilla.
Thus, similar to restricted positivity, the notion of restricted complete positivity for this qubit semigroup seems natural. In the Gaussian case such a concept was easily at hand due to the availability of the initial slip map, see \eqref{eq:rest_cpt_domain}. In this general semigroup case of a qubit, however, the character of the complete positivity domain $c{\mathcal P}$ requires further studies.

\section*{Conclusions}
\label{sec:conclusions}

In this article we investigate semigroup dynamics which are not of GKSL form and discuss their relevance for physical dynamics. Such dynamics arises naturally when considering the well-known master equation of quantum Brownian motion \eqref{eq:qbm_model}. The physical viability of a master equation is often connected to the complete positivity condition on the generator. However, demanding the complete positivity domain to be the full set of all quantum states is too restrictive as soon as initial system-environment correlations are taken into account, as witnessed in the context of dynamics with an initial slip.

In order to approach this problem, it is natural to discuss \emph{restricted (complete) positivity} of semigroup dynamics. We do that through the study a family of qubit master equations and discuss the parameter choices leading to completely positive, positive and restricted positive dynamics. It turns out that different to Gaussian dynamics, for qubit dynamics it is possible to have positive dynamics which are not completely positive. Furthermore, parameter choices leading to positivity violation can be characterized by the volume of the positivity domain, i.e.\ the volume of the set of initial states giving rise to positive dynamics for all times. The concept of restricted complete positivity is obviously way more challenging: while it can be formalized in the quantum Brownian motion case due to the explicit initial slip map, a similar characterization for the qubit dynamics is still missing and part of our future research.

\section*{Acknowledgements}
We thank Konstantin Beyer for helpful discussions and feedback on the manuscript. CB\ acknowledges support by the German Academic Scholarship Foundation.

\bibliography{literature}

\appendix

\section{Explicit expressions for the Gaussian map}\label{app:covariance_propagator} 

We find the time evolution of the covariance matrix \eqref{eq:covdynamicsmap} of the Gaussian state in the form of a map,
\begin{equation}\label{eq:covdynamicsmap2}
    \sigma(t) = X\sigma(0)X^T + Y,
\end{equation}
where
\begin{equation}\label{eq:cov_map_expressions}
    X=X(t) = \e^{Mt}\;\;\mbox{and}\;\; Y=Y(t)=\int_0^tds\; \e^{M(t-s)}(2D)\e^{M^T(t-s)}.
\end{equation}
We set $\eta=0$ in \eqref{eq:covdynamicsmatrices} such that
\begin{equation}\label{eq:covdynamicsmatrices2}
   M = \begin{pmatrix}
       0 & 1\\
       -\Omega^2 & -2\gamma
   \end{pmatrix} = A - \gamma\id,
\end{equation}
where $A^2=-\omega^2 \id$ with $\omega ^2 = \Omega^2-\gamma^2$ (assuming $\Omega>\gamma)$.
Thus,
\begin{equation}\label{eq:explicit_map}
    X(t) = \e^{Mt} = \e^{-\gamma t}\left(\id \cos\omega t +\frac{A}{\omega}\sin\omega t \right),
\end{equation}
which shows that $X(t)$ is stable for $\gamma>0$. Now the shift $Y(t)$ in \eqref{eq:covdynamicsmap2} is obtained easily by evaluating the time integral in \eqref{eq:cov_map_expressions} explicitly with expression \eqref{eq:explicit_map}.

The stationary state will have a covariance matrix given by
\begin{equation}\label{eq:stationarycov1}
    \sigma_\infty \equiv \lim_{t\rightarrow\infty} \sigma(t) = \lim_{t\rightarrow\infty} Y(t) = 2\int_0^\infty ds\; X(s) D  X^T(s),
\end{equation}
independent of the initial state. Again, the remaining time integral is evaluated with the explicit form \eqref{eq:explicit_map} of $X(s)$. An alternative expression for $\sigma_\infty$ is obtained from the stationary condition on \eqref{eq:covdynamics},
\begin{equation}\label{eq:stationarycov2}
    M\sigma_\infty + \sigma_\infty M^T + 2 D = 0.
\end{equation}

\section{Dekker-Valsakumar-condition and dynamics of the determinant}\label{app:determinant} 

We use 
\begin{equation}
\frac{d}{dt}\det\sigma(t) = \det\sigma(t)\,\mathrm{tr}\left[\sigma^{-1}(t)\dot\sigma(t)\right],
\end{equation}
and employ the time evolution of \eqref{eq:covdynamics},\begin{equation}\label{eq:covdynamics2}
    \dot\sigma = M\sigma + \sigma M^T + 2 D.
\end{equation}
From that we find
\begin{equation}
 \frac{d}{dt}\det\sigma(t) = 2\det\sigma(t)\,\left(\mathrm{tr} M +\mathrm{tr}[\sigma^{-1}D]\right).   
\end{equation}
For any positive, symmetric $2\times2$ matrix $P$, the arithmetic-geometric mean inequality reads $\mathrm{tr}P\ge2\sqrt{\det P}$. Choosing $P=\sigma^{-\frac{1}{2}}D\sigma^{-\frac{1}{2}}$, we can thus bound $\mathrm{tr}[\sigma^{-1} D] \ge 2\sqrt{\det D/\det \sigma}$. Evaluated at $t=t_0$ with $\det \sigma(t_0)=1/4$ and $\mathrm{tr}{M}=-2\gamma$, we find the desired inequality (\ref{eq:dekker_inequality}),
\begin{equation}\label{eq:dekker_inequality2}
    0> \frac{d}{dt}\left.\det\sigma(t)\right|_{t=t_0} \ge
    -\gamma+2\sqrt{\det D}.
\end{equation}

\section{Determining the positive region in the parameter space}\label{sec:appendixRP_P}

Against convention we parametrize the Bloch sphere $\vec{r}=\begin{pmatrix}
    \cos{\zeta}\\\cos{\eta}\sin{\zeta}\\ \sin{\eta}\sin{\zeta}
\end{pmatrix}$ for easier calculation. Here, $\zeta\in [0,\pi)$ is the polar angle between the radial line and the polar axis $\vec{x}$ and $\eta\in[0,2\pi)$ is the azimuthal angle.
Plugging this into \eqref{eq:DirectionVelocityVector} defines the function
\begin{equation}
    f_{\alpha,\beta}(\eta,\zeta)=\vec{r}^T G \vec{r}+\vec{r}\vec{c}=-\alpha \cos^2{\zeta}- \sin^2{\zeta}+\beta  \sin{\zeta} \sin{\eta}.
\end{equation}
The maximum of this function $f$ determines whether the dynamics is positive or non-positive:
\begin{equation}
    F(\alpha,\beta)\equiv\max_{\eta_{\text{max}},\zeta_{\text{max}}}f_{\alpha,\beta}(\eta,\zeta)=
    \begin{cases}
    >0,\text{the dynamics is non-positive.}\\
    \leq0,\text{the dynamics is positive.}
    \end{cases}
\end{equation} 
Thus, $F(\alpha,\beta)=0$ describes the boundary between positive and non-positive dynamics in the parameter plane $(\beta,\alpha)$.

First, we find the critical points $P_{\text{crit}}=(\eta_{\text{crit}},\zeta_{\text{crit}})$ by considering the first derivative:
\begin{subequations}
    \begin{align}
    0&\overset{!}{=}\left.\frac{\partial f}{\partial \zeta}\right|_{P_\text{crit}}= \cos{\zeta}\big(2(\alpha-1)\sin{\zeta}+\beta\sin{\eta}\big),\\
    &\Rightarrow\begin{cases}
        \text{(\RomanNumeralCaps{1})}& 2(\alpha-1)\sin{\zeta}+\beta\sin{\eta}=0,\\
         \text{(\RomanNumeralCaps{2})}& \cos{\zeta}=0,
    \end{cases}\\
    0&\overset{!}{=}\left.\frac{\partial f}{\partial \eta}\right|_{P_\text{crit}}= \beta \cos{\eta}\sin{\zeta},\\
    &\Rightarrow\begin{cases}
        \text{(\romannumeral1)}&\cos{\eta}=0,\\
         \text{(\romannumeral2)}&\sin{\zeta}=0.
    \end{cases}
\end{align}
\end{subequations}
Next, we determine whether the critical points correspond to local maxima. To this end, we evaluate the Hessian determinant
\begin{equation*}
    D=\left.\left[\frac{\partial^2 f}{\partial \zeta^2} \frac{\partial^2 f}{\partial \eta^2}-\left(\frac{\partial^2 f}{\partial \zeta \partial \eta}\right)^2\right]\right|_{P_\text{crit}}. 
\end{equation*}According to the second derivative test, there is a local maximum at a critical point, if $D>0$ and $\left.\left[\frac{\partial^2 f}{\partial \zeta^2}\right]\right|_{P_\text{crit}}<0$.

\paragraph{Case \RomanNumeralCaps{1}\romannumeral1}: Combining the conditions, we obtain for the critical point:
\begin{align*}
    (\eta_{\text{crit}1},\zeta_{\text{crit}1})&=\left(\pi/2,\arcsin{\left(-\frac{\beta}{2(\alpha-1)}\right)}\right),\\
    (\eta_{\text{crit}2},\zeta_{\text{crit}2})&=\left(3\pi/2,\arcsin{\left(\frac{\beta}{2(\alpha-1)}\right)}\right).
\end{align*} 
These solutions do not exist for all values of $(\alpha,\beta)$. In particular, they are not valid in the region of the parameter plane bounded by the curves
 $\alpha(\beta)=\begin{cases}
    1+\frac{\beta}{2}, & \beta \in [-1,1]\\
    1-\frac{\beta}{2}, & \beta \in [-1,1]
\end{cases}$ because, in this region, $\left| \frac{\beta}{2(\alpha-1)}\right|>1$ and the argument of the inverse sine lies outside its domain. Beneath this region, the Hessian test instead identifies a local maximum. 
\begin{figure}[h!]
    \centering
    \includegraphics[width=0.9\linewidth]{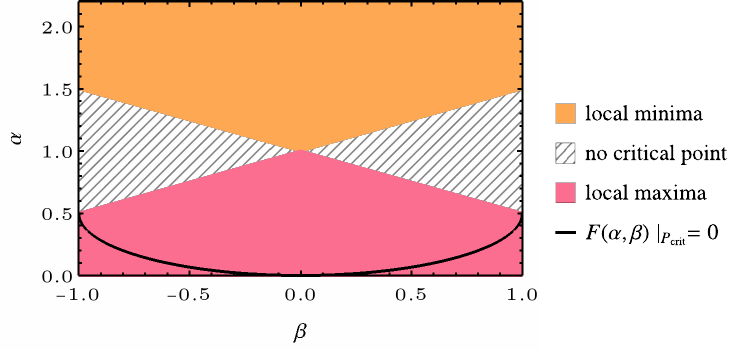}
    \caption{Parameter regions in the $(\beta,\alpha)$-plane in which the critical points exist are colored. In the hatched region, the inverse sine is undefined, and therefore no critical points exist. In the upper region, the Hessian test identifies the critical points as local minima of $f$, whereas in the lower region they correspond to local maxima. The black curve is given by \eqref{eq:BoundaryPRP}.}
    \label{fig:ParamplaneMinMax}
\end{figure}

Within this maximum region, the boundary between positive and non-positive dynamics is determined by the condition $F(\alpha,\beta)\big|_{P_\text{crit}}=0$, which reads 
\begin{equation}\label{eq:BoundaryPRP2}
    \alpha(\beta)=\frac{1}{2}\left(1-\sqrt{1-\beta^2}\right).
\end{equation}
All dynamics beneath this curve are non-positive and above this curve are positive. Note, that completely positive dynamics are also positive. These findings are summarized in Fig. \ref{fig:ParamplaneMinMax}.

\paragraph{Case \RomanNumeralCaps{1}\romannumeral2}: Combining the conditions, we obtain four critical points:
\begin{align*}
    \eta_{\text{crit}1,2}=0, \pi,\\
    \zeta_{\text{crit}1,2}=0, \pi. 
\end{align*}
We shorten the discussion by observing, that $F(\alpha,\beta)=-\alpha <0$ for all four critical points. Hence, even in the case that there exists a local maximum at those critical points, the velocity vector of the pure states is pointing inwards and the resulting dynamics is positive. 

\paragraph{Case \RomanNumeralCaps{2}\romannumeral1}: Combining the conditions, we obtain four critical points: 
\begin{align*}
    (\eta_{\text{crit}1,2},\zeta_{\text{crit}1,2})&=\left(\pm \frac{\pi}{2},\pm\frac{\pi}{2}\right),\\
    (\eta_{\text{crit}3,4},\zeta_{\text{crit}3,4})&=\left(\pm \frac{\pi}{2},\mp\frac{\pi}{2}\right).
\end{align*}
Once again, we find that the boundary function is negative at all four critical points. Specifically, $F(\alpha,\beta)\big|_{P_{\text{crit}1,2}}= -1+\beta$  and $F(\alpha,\beta)\big|_{P_{\text{crit}3,4}}= -1-\beta$. Since $\beta\in(-1,1)$, both expressions are strictly negative. Therefore, even if these critical points correspond to local maxima, the velocity vector at the associated pure states points inward, toward the interior of the Bloch ball. Consequently, the resulting dynamics are positive. 

\paragraph{Case \RomanNumeralCaps{2}\romannumeral2}: These two cases are mutually exclusive. Hence, they do not result in a critical point.

\section{Existence of a set of initial states leading to positive dynamics}
\label{sec:appendix_tetrahedron}
Suppose two initial conditions $\rho_1$, $\rho_2$ give rise to positive dynamics \linebreak$\mathcal{E}_t \rho_i \in \mathcal{S} \ \forall t$, then all convex combinations lead to physical dynamics and $\mathcal{E}_t \left[ \lambda \rho_1 + (1 - \lambda) \rho_2 \right] \in \mathcal{S}$ holds for all $t$ and $\lambda \in [0,1]$.
In the case of the qubit discussed here, it is sufficient to identify four distinct, non-coplanar points within the Bloch ball (which form a tetrahedron) that give rise to physical dynamics. Therefore, all states that are convex combinations of these points (i. e., those contained within the volume of the tetrahedron) give rise to physical dynamics.

Under the map \eqref{eq:Blochmap}, the norm of the time-evolved edges of the tetrahedron must never exceed one:
\begin{equation}
    \text{for all }t: 1\geq \left|\vec{r}^\prime\right|=\sqrt{|x|^2 \e^{-2\alpha t} +|y|^2 \e^{-2t} + |\beta(1-\e^{-t})+z \e^{-t}|^2}.
\end{equation}
For $|\beta|<1$, we can always find an $\epsilon>0$ such that $|\beta|+\epsilon\leq1$. 
We identify the edges of the tetrahedron as $\vec{r}_{1,2}=(0,0,\beta\pm \epsilon), \vec{r}_3=(\epsilon,0,\beta)$ and $\vec{r}_4=(0,\epsilon,\beta)$. These points are not coplanar as $\vec{r}_1,\vec{r}_2$ and $\vec{r}_3$ lie within the plane $y=0$, while $\vec{r}_4$ does not.
We can asses the norm of the time-evolved state vectors $\vec{r}_i$ under the evolution by the map \eqref{eq:Blochmap} for each of the states individually
\begin{subequations}
	\begin{align}
		&i=1,2: &&\left|\beta \pm \epsilon\ e^{-t_0}\right| \leq \left|\beta\right|  + \epsilon  \leq 1, \\
		&i=3: &&\sqrt{ \epsilon ^2 \ e^{-2 \alpha t_0}+\left|\beta\right| ^2} \leq \sqrt{\epsilon ^2 \ +\left|\beta\right| ^2} \overset{\text{\tiny{triangle inequality}}}{\leq} |\beta|+\epsilon \leq 1,\\
		&i=4: &&\sqrt{ \epsilon ^2 \ e^{-2 t_0}+\left|\beta\right| ^2} \leq \sqrt{ \epsilon ^2 \ +\left|\beta\right| ^2} \overset{\text{\tiny{triangle inequality}}}{\leq} |\beta|+\epsilon \leq 1.
	\end{align}
\end{subequations}
Hence, the map is positive for those four initial states corresponding to $\{\vec{r}\}_{i=1}^4$. All qubit states represented by Bloch vectors enclosed in the formed tetrahedron lead to positive dynamics under the map as well. 
\end{document}